\documentstyle[aps,prb,multicol,epsf]{revtex}

\begin{document}
\draft

\title{Charge-orbital ordering and phase-separation
  in the two-orbital model for manganites: \\
  Roles of Jahn-Teller phononic and Coulombic interactions}

\author{Takashi Hotta}
\address{Institute for Solid State Physics, University of Tokyo,
  7-22-1 Roppongi, Minato-ku, Tokyo 106-8666, Japan}

\author{Andre Luiz Malvezzi}
\address{Departamento de F\'\i sica - Faculdade de Ci\^encias -
  Universidade Estadual Paulista \\ Caixa Postal 473, 17.033-360,
  Bauru, SP, Brazil}

\author{Elbio Dagotto}
\address{Department of Physics and National High Magnetic Field
  Laboratory, Florida State University, Tallahassee, FL 32306}

\date{\today}

\maketitle

\begin{abstract}
The main properties of realistic models for manganites are studied
using analytic mean-field approximations and computational numerical
methods, focusing on the two-orbital model with electrons interacting
through Jahn-Teller (JT) phonons and/or Coulombic repulsions. 
Analyzing the model including both interactions by the combination 
of the mean-field approximation and the exact diagonalization method,
it is argued that the spin-charge-orbital structure in the
insulating phase of the purely JT-phononic model is not qualitatively
changed by the inclusion of the Coulomb interactions.
As an important application of the present mean-field approximation,
the $CE$-type AFM state, the charge-stacked structure along the
$z$-axis, and the $(3x^2-r^2)$/$(3y^2-r^2)$-type orbital ordering are 
successfully reproduced based on the JT-phononic model for the
half-doped manganite, in agreement with recent Monte Carlo simulation
studies by Yunoki {\it et al.}
Topological arguments and the relevance of the Heisenberg
exchange among localized $t_{\rm 2g}$ spins explains why the
inclusion of the nearest-neighbor Coulomb interaction does not destroy 
the charge stacking structure.
It is also verified that the phase-separation tendency is observed
both in purely JT-phononic and purely Coulombic models in the vicinity
of the hole undoped region, as long as realistic hopping matrices are
used. This highlights the qualitative similarities of both approaches,
and the relevance of mixed-phase tendencies in the context of
manganites.
In addition, the rich and complex phase diagram of the two-orbital
Coulombic model in one dimension is presented. 
Our results provide robust evidence that Coulombic and JT-phononic
approaches to manganites are not qualitatively different ways to carry
out theoretical calculations, but they share a variety of common
features. 
\end{abstract}


\begin{multicols}{2}
\narrowtext

\section{Introduction}

The recent discovery of the colossal magnetoresistance (CMR)
phenomenon in manganese oxides has triggered a huge experimental and 
theoretical effort to understand its origin.\cite{jin,tokura}
The large changes in resistivity observed in experiments usually
involve both a low-temperature or high magnetic field ferromagnetic
(FM) metallic phase, and a high-temperature or low magnetic field
insulating phase.
For this reason, to address the CMR effect, it appears unavoidable to
have a proper understanding of the phases competing with the metallic
FM regime, which as a first approximation can be rationalized based on 
the standard double exchange (DE) mechanism.\cite{zener,gennes}
In fact, the phase diagram of manganites has revealed a very complex
structure, clearly showing that metallic ferromagnetism is just one of
the several spin, orbital, and charge arrangements that are possible
in the manganese oxides.\cite{phase,cheong}

The theoretical study of manganites started decades ago when the DE
ideas to explain the FM-phase were proposed.
Unfortunately, the model used in those early calculations involved
only one orbital, and the many-body techniques used in its analysis 
were mostly restricted to crude mean-field approximations (MFA).
Quite recently, the first computational studies of the one-orbital
model for manganites have been presented by Yunoki {\it et al.}
\cite{yunoki,yunoki2,yunoki3}
While the expected FM-phase emerged clearly from such analysis,
several novel features were identified, notably phase-separation (PS)
tendencies close to $n$=1 (where $n$ is the $e_{\rm g}$ electron
number density per site) between an antiferromagnetic (AFM) insulating
phase and a metallic FM state.
Several calculations have confirmed these tendencies toward
mixed-phase characteristics, using a variety of
techniques.\cite{guinea}
Moreover, the computational work has been extended to include
non-cooperative Jahn-Teller (JT) phonons\cite{yunoki4} and PS
tendencies involving spin-FM phases that differ in their orbital
arrangement, i.e., staggered vs. uniform, were again identified using
one-dimensional (1D) and two-dimensional (2D) clusters.
This illustrates the relevance of the orbital degree of freedom in the 
analysis of manganites, and confirms the importance of mixed-phase
characteristics in this context.
The theoretical, mostly computational, calculations have been recently 
summarized by Moreo {\it et al.},\cite{moreo} where it has been argued
that a state with clusters of one phase embedded into another should
have a large resistivity and a large compressibility.
Other predictions in mixed-phase regimes involve the presence of a
robust pseudogap in the density of states,\cite{moreo2}
similar to the results obtained using angle-resolved photoemission
experiments applied to bilayer manganites.\cite{dessau}
A large number of experiments \cite{moreo} have reported the presence
of inhomogeneities in real manganites,\cite{teresa,mydosh,cheong2}
results compatible with those obtained using computational studies.
More recently, the influence of disorder on metal-insulator
transitions of manganite models that would be of first-order character
without disorder has been proposed\cite{disorder} 
as the origin of the large cluster coexistence reported in experiments
for several compounds.\cite{mydosh,cheong2}  
It is clear that intrinsic tendencies toward inhomogeneous states is
at the heart of manganite physics, and simple scenarios involving
small polarons are not sufficient to understand the mixed-phase
tendencies observed experimentally.

Although the presence of mixed-phase tendencies competing with the 
FM-phase is by now well-established and a considerable progress
has been achieved in the theoretical study of models for manganites,
many issues still remain to be investigated in this context.
One of them is related with the notorious charge ordering phenomenon,
characteristic of the narrow-band manganites.
Especially at $n$=0.5, the so-called $CE$-type AFM phase has been 
widely observed in materials such as La$_{0.5}$Ca$_{0.5}$MnO$_3$ and
Nd$_{0.5}$Sr$_{0.5}$MnO$_3$.
In spite of the considerable importance of this phase in view of its
relevance for the CMR effect reported at high hole
density,\cite{tokura} it is only recently that the $CE$-phase has 
been theoretically understood using the analytic MFA in the JT-phononic
model \cite{hotta,hotta2} and numerical computational 
techniques.\cite{hotta3,hotta4} 
The importance of the zigzag chains of the $CE$-state, along which the
spins are parallel, has been remarked in these recent efforts.
In fact, it has been shown that even without electron-phonon or
Coulomb interactions the $CE$-state structure is nevertheless stable,
and in this limit it arises from a simple band-insulator picture.
Previous work, such as the pioneering results of
Goodenough,\cite{goodenough} were based on the $assumption$ of a
checkerboard charge state upon which the spin and orbital arrangements
were properly calculated, but they did not consider the competition
with other charge disordered phases in a fully unbiased calculation as 
carried out in Refs.\onlinecite{hotta,hotta2,hotta3,hotta4}.

Here a particular feature of the CO $CE$-phase structure should be
remarked.
In the $x$-$y$ plane, the Mn$^{3+}$ and Mn$^{4+}$ ions arrange in a 
checkerboard pattern, which naively seems to arise quite naturally
from the presence of a strong long-range Coulomb
interaction.\cite{goodenough}
However, contrary to this naive expectation, the same CO pattern
stacks in experiments along the $z$-axis without any change, although
the $t_{\rm 2g}$ spin direction alternates from plane to plane.
In other words, the Mn$^{3+}$ and Mn$^{4+}$ ions in the $CE$-type 
AFM structure do $not$ form a three-dimensional (3D) Wigner-crystal
structure, clearly suggesting that the origin of the CO phase in the
half-doped manganite $cannot$ be caused exclusively by a dominant
long-range Coulomb interaction. 
Other interactions should be important as well.
In this work, it is shown that the charge-stacked (CS) CO-state 
in the $CE$-type structure found in experiments can also be obtained
in the simple MFA for the purely JT-phononic model.
In addition, it is shown that this structure is not easily destroyed
by the inclusion of the nearest-neighbor Coulomb interaction. 
The exchange $J_{\rm AF}$ between the $t_{\rm 2g}$ spins plays a key
role in the stabilization of the CS structure.
The details of the calculation are discussed here. 
The present semi-analytical results are in excellent agreement with
recent Monte Carlo (MC) simulations\cite{hotta4} and provide a simple
formalism to rationalize these computational results.

On the other hand, it is possible to take another approach to
understand the CO state in manganites including the $CE$-type
structure, by emphasizing the role of the on-site Coulomb
interaction.\cite{Coulomb} 
In this direction of study, 
several results are available in the literature.
For instance, Ishihara {\it et al.} discussed the spin-charge-orbital
ordering from the limit of infinite strong electronic correlation,
namely, under the approximation of no double
occupancy.\cite{ishihara}
Maezono {\it et al.} provided information on the overall features of
the phase diagram for manganites by using a mean-field
calculation.\cite{maezono}
These two mean-field approaches do not reproduce the $CE$-state of
half-doped manganites.
Mizokawa and Fujimori discussed the stabilization of the $CE$-type
structure by using the model Hartree-Fock approximation.\cite{mizokawa}
However, to the best of our knowledge, the origin of the
CS structure in the $CE$-type state has not been clarified based on the 
Coulombic scenario, 
although several properties of manganites have been reproduced.
In this paper, a possible way to understand the $CE$-type state
with the CS structure in the purely Coulombic model even without
the long-range repulsion is briefly discussed.

The present paper contains information about other subjects as well.
In spite of the importance of additional interactions besides the 
long-range Coulombic one to stabilize the proper charge-stacked
$n$=$0.5$ state, as discussed before,
it is clear that at least the on-site Coulomb interactions are 
very strong and their influence should be considered in
realistic calculations.
Most of the previous analytic and computational works that reported 
the PS tendency and the CO phase have used electrons interacting
among themselves indirectly through their coupling with localized  
$t_{\rm 2g}$ spins or with JT phonons.
The JT-phononic model has been extensively studied by the present
authors since the explicit inclusion of on-site Coulomb interactions
diminish dramatically the feasibility of the computational work.
In addition, it has been found that the JT-phononic model explains quite
well several experimental results, even if the Coulomb interaction is
not included explicitly.
Moreover, it should be noted that the energy gain due to the static JT
distortion is maximized when only one $e_{\rm g}$ electron is located
at the JT center.
Confirming this expectation, the double-occupancy of a given orbital
has been found to be negligible in previous investigations at
intermediate and large values of the electron-phonon coupling, and in 
this situation, adding an on-site repulsion would not alter the
physics of the state under investigation.
Nevertheless, although the statements above are believed to be
qualitatively correct, it should be checked explicitly in a more
realistic model that indeed the physics found in the extreme case of
only phononic interactions survives when both JT-phononic and
Coulombic interactions are considered.
Thus, another purpose of this paper is to clarify the validity of the
JT-phononic only model by showing that the inclusion of the Coulomb
interaction does  not bring qualitative changes in the conclusions
obtained from the purely phononic model.
For this purpose, an analytic-numeric combination is employed, namely,
the MFA including Coulombic terms is developed in detail and the
results are compared with exact diagonalization (ED) data for a small
cluster.
In addition, the PS tendency is reexamined briefly within the
framework of MFA, and some comments on previous results are provided.
It is clear that addressing whether purely JT phononic or purely
Coulombic approaches lead or not to qualitatively different phase 
diagrams is an important area of investigation.

As a special case of the general goal described in the previous
paragraph, the issue of PS in the presence of Coulomb interactions
will also be studied in this paper using computational techniques. 
In fact, it has not been analyzed whether models with two orbitals and 
Coulomb repulsions, i.e. without JT-phonons, lead to mixed-phase
tendencies as pure phononic models do.
In this paper it is reported that a study of a simple one-dimensional
two-orbital model without phonons but with strong Coulomb
interactions, produces PS tendencies similarly as reported in previous
investigations by our group.\cite{yunoki4}
Several features are in qualitative agreement with those observed
using JT-phonons, illustrating the similarities between the purely
phononic and the purely Coulombic approaches to manganites, and the
clear relevance of mixed-phase characteristics for a proper
description of these compounds.
It is concluded that simple models for manganites, once studied with
robust computational techniques, are able to reveal a complex phase
diagram with several phases having characteristics quite similar to
those observed experimentally.

The organization of the paper is as follows.
In Sec.~II, a theoretical model for manganites is introduced, and
several reduced Hamiltonians are discussed in preparation for the
subsequent investigations. 
Section III is devoted to the MFA.
The formulation for the MFA is provided and a comparison with the ED
result is shown to discuss the validity of the MFA.
It is argued that the JT-phononic interaction plays a role 
more relevant than the Coulombic interaction as suggested from the
viewpoint of the continuity of the orbital-ordered (OO) phase.
In Sec.~IV, as an application of the present MFA, the CO/OO structure
in half-doped manganites is studied.
Especially, the origin of the $CE$-type state with the CS structure 
is discussed in detail.
In Sec.~V, the PS tendency is studied in the framework
of the MFA for the purely JT-phononic and Coulombic models.
In Sec.~VI, the two-orbital 1D Hubbard model is analyzed using the
density matrix renormalization group (DMRG) technique. 
Even in the purely Coulombic model, PS tendencies and CO phases are
observed.
Finally in Sec.~VII, the main results of the paper are summarized.
The main overall conclusion is that studies based upon purely 
JT-phononic or Coulombic approaches do $not$ differ substantially
in their results, and there is no fundamental difference between them.
For reasons to be described below, of the two approaches the best
appears to be the phononic one due to its ability to select the proper
orbital ordering pattern uniquely.
PS appears in purely phononic and purely Coulombic approaches as well.
In Appendix A, the relation between the PS scenarios for cuprates and 
manganites is discussed.
Throughout this paper, units such that $\hbar$=$k_{\rm B}$=1 are used.

\section{Models for Manganites}

In order to explain the existence of ferromagnetism at low
temperatures in the manganites, the DE mechanism has been usually
invoked.\cite{zener,gennes}
In this framework, holes can improve their kinetic energy by
polarizing the spins in their vicinity.
If only one $e_{\rm g}$ orbital is used, the FM Kondo or one-orbital
model is obtained,\cite{furukawa} and several results for this model
are already available in the literature.
\cite{yunoki,yunoki2,yunoki3}
Although the important PS tendencies have been identified in this
context, it is clear that the one-orbital model is not sufficient to
describe the rich physics of the manganites, where the two active
orbitals play a key role and the CO state close to density 0.5 is
crucial for the appearance of a huge MR effect at this density.
Thus, in this paper, the two-orbital model is discussed to understand
the manganite physics, by highlighting the complementary roles of the
JT phonon and Coulomb interactions.

\subsection{Hamiltonian}

Let us consider doubly-degenerated $e_{\rm g}$ electrons, tightly
coupled to localized $t_{\rm 2g}$ spins and local distortions
of the MnO$_6$ octahedra.
This situation is described by the Hamiltonian $H$ composed of five
terms 
\begin{equation}
  \label{Hamiltonian}
  H = H_{\rm kin} + H_{\rm Hund} + H_{\rm AFM} 
  + H_{\rm el-ph} + H_{\rm el-el}.
\end{equation}
The first term indicates the hopping motion of 
$e_{\rm g}$ electrons, given by,
\begin{eqnarray}
  H_{\rm kin} =-\sum_{{\bf ia}\gamma \gamma'\sigma}
  t^{\bf a}_{\gamma \gamma'} d_{{\bf i} \gamma \sigma}^{\dag}
  d_{{\bf i+a} \gamma' \sigma},
\end{eqnarray}
where $d_{{\bf i}{\rm a}\sigma}$ ($d_{{\bf i}{\rm b}\sigma}$) is the
annihilation operator for an $e_{\rm g}$-electron with spin $\sigma$
in the $d_{x^2-y^2}$ ($d_{3z^2-r^2}$) orbital at site ${\bf i}$,
${\bf a}$ is the vector connecting nearest-neighbor sites,
and $t^{\bf a}_{\gamma \gamma'}$ is the nearest-neighbor hopping
amplitude between $\gamma$- and $\gamma'$-orbitals along the 
${\bf a}$-direction.
The amplitudes are evaluated from the overlap integral between manganese
and oxygen ions,\cite{Slater} given by
\begin{eqnarray}
  t_{\rm aa}^{\bf x}
  =-\sqrt{3}t_{\rm ab}^{\bf x}
  =-\sqrt{3}t_{\rm ba}^{\bf x}
  =3t_{\rm bb}^{\bf x}=t,
\end{eqnarray}
for the $x$-direction,
\begin{eqnarray}
  t_{\rm aa}^{\bf y}
  =\sqrt{3}t_{\rm ab}^{\bf y}
  =\sqrt{3}t_{\rm ba}^{\bf y}
  =3t_{\rm bb}^{\bf y}=t,
\end{eqnarray}
for the $y$-direction, and 
\begin{eqnarray}
  t_{\rm bb}^{\bf z}=4t/3,
  t_{\rm aa}^{\bf z}=t_{\rm ab}^{\bf z}=t_{\rm ba}^{\bf z}=0,
\end{eqnarray}
for the $z$-direction.
Note that $t_{\rm aa}^{\bf x}$ (equal to $t_{\rm aa}^{\bf y}$ by
symmetry) is taken as the energy scale $t$.

The second term is the Hund coupling between the $e_{\rm g}$ electron
spin ${\bf s}_{\bf i}$ and the localized $t_{\rm 2g}$ spin 
${\bf S}_{\bf i}$, given by
\begin{equation}
  H_{\rm Hund} = -J_{\rm H} \sum_{\bf i}
  {\bf s}_{\bf i} \cdot {\bf S}_{\bf j},
\end{equation}
with 
${\bf s}_{\bf i}=
\sum_{\gamma\alpha\beta}d^{\dag}_{{\bf i}\gamma\alpha}
\bbox{\sigma}_{\alpha\beta}d_{{\bf i}\gamma\beta}$,
where $J_{\rm H}$($>$0) is the Hund coupling, and 
$\bbox{\sigma}$=$(\sigma_x, \sigma_y, \sigma_z)$ are the Pauli
matrices.
Note here that the $t_{\rm 2g}$ spins are assumed classical and
normalized to $|{\bf S}_{\bf i}|$=1.

The third term accounts for the $G$-type AFM property of manganites in
the fully hole doped limit, given by
\begin{equation}
  H_{\rm AFM} = J_{\rm AF} \sum_{\langle {\bf i,j} \rangle}
  {\bf S}_{\bf i} \cdot {\bf S}_{\bf j},
\end{equation}
where $J_{\rm AF}$ is the AFM coupling between nearest neighbor 
$t_{\rm 2g}$ spins.
The value of $J_{\rm AF}$ will be small in units of $t$
to make it compatible with
experiments for the fully hole-doped CaMnO$_3$ compound.\cite{perring} 

In the fourth term, the coupling of $e_{\rm g}$ electrons to the
lattice distortion is considered as\cite{kanamori,millis,allen}
\begin{eqnarray}
  && H_{\rm el-ph} = g \sum_{{\bf i}\sigma}
  [Q_{1{\bf i}} (d_{{\bf i} a\sigma}^{\dag}
  d_{{\bf i}a\sigma}+d_{{\bf i} b\sigma}^{\dag}d_{{\bf i}b\sigma})
  \nonumber \\ 
  &+& Q_{2{\bf i}} (d_{{\bf i} a\sigma}^{\dag}
  d_{{\bf i}b\sigma}+d_{{\bf i} b\sigma}^{\dag}d_{{\bf i}a\sigma})
  + Q_{3{\bf i}} (d_{{\bf i} a\sigma}^{\dag}d_{{\bf i}a\sigma}
  -d_{{\bf i} b\sigma}^{\dag}d_{{\bf i}b\sigma})]
  \nonumber \\
  &+& (1/2) \sum_{\bf i} [k_{\rm br}Q_{1{\bf i}}^2
  +k_{\rm JT}(Q_{2{\bf i}}^2+Q_{3{\bf i}}^2)],
\end{eqnarray}
where $g$ is the coupling constant between $e_{\rm g}$ electrons
and distortions of the MnO$_6$ octahedron,
$Q_{1{\bf i}}$ is the breathing-mode distortion,
$Q_{2{\bf i}}$ and $Q_{3{\bf i}}$ are, respectively, the JT
distortions for the $(x^2-y^2)$- and $(3z^2-r^2)$-type modes,
and $k_{\rm br}$ ($k_{\rm JT}$) is the spring constant for 
the breathing (JT) mode distortions.
Here the distortions are treated adiabatically, but
the validity of this treatment should be discussed.
In general, the adiabaticity is judged from the ratio $\eta$=
$\tau_{\rm e}/\tau_{\rm i}$,
where $\tau_{\rm e}$ and $\tau_{\rm i}$ are characteristic time
scales for the motion of electrons and ions, respectively.
If $\eta$ is much less than unity, the adiabatic approximation
holds, since the motion of ions is very slow compared to that of
electrons.
Due to the relations $\tau_{\rm i}$$\sim$$\omega^{-1}$ and
$\tau_{\rm e}$$\sim$$t^{-1}$, where $\omega$ is the frequency of the 
lattice distortion, $\eta$$\sim$$\omega/t$ is obtained.
In the manganite, $t$ is 0.2-0.5 eV,\cite{hopping} 
while $\omega$ is about $500$-$600$cm$^{-1}$.\cite{Iliev}
Thus, the adiabatic approximation is in principle valid in the study 
of the JT distortion of manganites.
An important energy scale characteristic of $H_{\rm el-ph}$ is the
static JT energy, defined by
\begin{eqnarray}
  E_{\rm JT}=g^2/(2k_{\rm JT}),
\end{eqnarray}
which is naturally obtained by the scaling of 
$Q_{\mu{\bf i}}=(g/k_{\rm JT})q_{\mu{\bf i}}$
with $\mu$=1,2, and 3, where 
$g/k_{\rm JT}$ is the typical length scale for the JT distortion.
By using $E_{\rm JT}$ and $t$, it is convenient to introduce the 
non-dimensional electron-phonon coupling constant $\lambda$ as
\begin{eqnarray}
  \lambda = \sqrt{2E_{\rm JT}/t}.
\end{eqnarray}
The characteristic energy for the breathing-mode distortion
is given by $E_{\rm br}$=$g^2/(2k_{\rm br})$=$E_{\rm JT}/\beta$
with $\beta$=$k_{\rm br}/k_{\rm JT}$.

The last term indicates the Coulomb interactions between 
$e_{\rm g}$ electrons, expressed by 
\begin{eqnarray}
  H_{\rm el-el} &=& U \sum_{{\bf i}\gamma} 
  \rho_{{\bf i}\gamma\uparrow} \rho_{{\bf i}\gamma\downarrow}
  +U' \sum_{{\bf i}\sigma \sigma'} 
  \rho_{{\bf i}{\rm a}\sigma} \rho_{{\bf i}{\rm b}\sigma'} \nonumber
  \\ &+&J \sum_{{\bf i}\sigma \sigma'}
  d^{\dag}_{{\bf i}{\rm a}\sigma}d^{\dag}_{{\bf i}{\rm b}\sigma'}
  d_{{\bf i}{\rm a}\sigma'}d_{{\bf i}{\rm b}\sigma} 
  + V \sum_{\langle {\bf i,j} \rangle}
  \rho_{\bf i} \rho_{\bf j},
\end{eqnarray}
with $\rho_{{\bf i} \gamma\sigma}$=
$d_{{\bf i} \gamma \sigma}^{\dag}d_{{\bf i} \gamma \sigma}$ and 
$\rho_{\bf i}$=$\sum_{\gamma \sigma}\rho_{{\bf i} \gamma\sigma}$,
where $U$ is the intra-orbital Coulomb interaction, $U'$ is the
inter-orbital Coulomb interaction, $J$ is the inter-orbital exchange
interaction, and $V$ is the nearest-neighbor Coulomb interaction.
Note that the Hund-coupling term $H_{\rm Hund}$ between conduction and
localized spins also arises from Coulombic effects, but in this paper,
the ``Coulomb interaction'' refers to the direct electrostatic
repulsion between $e_{\rm g}$ electrons. 
Note also that the relation $U$=$U'$+$2J$ holds in the localized ion
system.
Although the validity of this relation may not be guaranteed in the
actual material, it is assumed to hold in this paper to reduce the
parameter space in the calculation.

Finally, a comment about the cooperative JT effect which is not
included in the phononic models is discussed here.
In principle, a Heisenberg-like coupling between nearest-neighbor 
phonons should also be added to account for the fact that when 
a given octahedra is elongated in one direction, its neighbors 
deform in an opposite way.\cite{kanamori}
However, this cooperative effect has not been studied in detail using 
computational techniques and here the focus of the paper will be on
results obtained with non-cooperative phonons.
The rich phase diagram and good qualitative agreement with experiments 
justifies {\it a-posteriori} this extra approximation.
In fact, in the undoped limit, there are no essential differences 
between the results with and without the cooperative
effect.\cite{hotta5} 

\subsection{Reduced Hamiltonians}

The Hamiltonian Eq.~(\ref{Hamiltonian}) is believed to define an
appropriate model for manganites, but it is quite difficult to solve
it exactly. 
In order to investigate further the properties of manganites based on
$H$, some simplifications and approximations are needed.
A simplification without the loss of essential physics is to take the 
widely used limit $J_{\rm H}$=$\infty$. 
In such a limit, the $e_{\rm g}$ electron spin perfectly aligns along
the $t_{\rm 2g}$ spin direction, reducing the number of degrees of
freedom.
Moreover, there is a clear advantage that both the intra-orbital
on-site term and the exchange term can be neglected in 
$H_{\rm el-el}$. Thus, the following simplified model is obtained:
\begin{eqnarray}
  && H^{\infty} = -\sum_{{\bf ia}\gamma \gamma'}
  {\cal S}_{\bf i,i+a} t^{\bf a}_{\gamma \gamma'} 
  c_{{\bf i} \gamma}^{\dag}c_{{\bf i+a} \gamma'}
  + J_{\rm AF} \sum_{\langle {\bf i,j} \rangle}
  {\bf S}_{\bf i} \cdot {\bf S}_{\bf j} \nonumber \\ 
  &+& E_{\rm JT} \sum_{\bf i}
  [2( q_{1{\bf i}} n_{\bf i} 
  + q_{2{\bf i}} \tau_{x{\bf i}}
  + q_{3{\bf i}} \tau_{z{\bf i}})
  + \beta q_{1{\bf i}}^2 + q_{2{\bf i}}^2 +q_{3{\bf i}}^2] 
  \nonumber  \\
  &+& U' \sum_{\bf i} n_{{\bf i}{\rm a}} n_{{\bf i}{\rm b}}
  + V \sum_{\langle {\bf i,j} \rangle} n_{\bf i}n_{\bf j},
\end{eqnarray}
where $c_{{\bf i} \gamma}$ is the spinless $e_{\rm g}$ electron
operator, given by\cite{muller} 
$c_{{\bf i} \gamma}$=
$\cos(\theta_{\bf i}/2)d_{{\bf i}\gamma \uparrow}$+
$\sin(\theta_{\bf i}/2)e^{-i\phi_{\bf i}}d_{{\bf i}\gamma\downarrow}$,
the angles $\theta_{\bf i}$ and $\phi_{\bf i}$ define the direction of
the classical $t_{\rm 2g}$ spin at site ${\bf i}$,
$n_{{\bf i} \gamma}$=$c_{{\bf i} \gamma}^{\dag}c_{{\bf i} \gamma}$,
$n_{\bf i}$=$\sum_{\gamma}n_{{\bf i} \gamma}$,
$\tau_{x\bf i}$=
$c_{{\bf i}a}^{\dag}c_{{\bf i}b}$+$c_{{\bf i}b}^{\dag}c_{{\bf i}a}$,
and $\tau_{z\bf i}$=
$c_{{\bf i}a}^{\dag}c_{{\bf i}a}$$-$$c_{{\bf i}b}^{\dag}c_{{\bf i}b}$.
Here ${\cal S}_{\bf i,j}$ denotes the change of hopping amplitude 
due to the difference in angles between $t_{\rm 2g}$ spins at 
sites ${\bf i}$ and ${\bf j}$, given by
\begin{eqnarray}
  {\cal S}_{\bf i,j} &=& \cos (\theta_{\bf i}/2)\cos (\theta_{\bf j}/2)
  \nonumber \\ 
  &+& \sin (\theta_{\bf i}/2)\sin (\theta_{\bf j}/2) 
  e^{-i(\phi_{\bf i}-\phi_{\bf j})}.
\end{eqnarray}
In principle, $\theta_{\bf i}$ and $\phi_{\bf i}$ should be optimized 
to provide the lowest-energy state.
However, in the following analytic approach,
only spin configurations such that nearest neighbor
$t_{\rm 2g}$ spins are either FM or AFM will be considered.
Namely, the spin canting state is excluded from the outset.
The validity of this simplification is later checked by comparison
with the numerical results obtained from a relaxation technique.
Note that this simplification does not restrict us to only fully FM or
$G$-type AFM states, but a wide variety of other states, such as 
the $CE$-type one, can also be studied.

Based on the reduced Hamiltonian $H^{\infty}$, it is possible to
consider two scenarios for manganites, as schematically shown in
Fig.~1.
Temporarily, $V$ will not be taken into account to focus on the
competition between $E_{\rm JT}$ and $U'$.
One of them is the JT-phonon scenario in which $U'$ is considered as a
correction into the purely JT-phononic model 
$H_{\rm JT}^{\infty}$=$H^{\infty}$($U'$=0,$V$=0) taken as starting
point.
Another is the Coulomb scenario where the JT distortion
is a correction into the purely Coulombic 
Hubbard-like model $H_{\rm C}^{\infty}$=$H^{\infty}$($E_{\rm JT}$=0).
Of course, if the many-body analysis were very accurate then there
should be no difference between the final conclusions obtained from
both scenarios when the realistic situation, $E_{\rm JT}$$>$0 and
$U'$$>$0, is approached.
However, if this realistic situation can be continuously obtained
from the limiting cases $U'$=0 or $E_{\rm JT}$=0, then it is enough
to consider only the limiting models $H_{\rm JT}^{\infty}$ or 
$H_{\rm C}^{\infty}$ to grasp the essence of the manganite physics.
On the other hand, it may also occur that starting in one of the two
extreme cases a singularity exists preventing a smooth continuation
into the realistic region of parameters. 
In this case, just one of the scenarios would be valid.
For investigations in this context, the simplified model $H^{\infty}$
certainly provides a good stage to examine the roles of $U'$ and
$E_{\rm JT}$. 
In the following section, this point will be discussed in detail.

\begin{figure}[h]
  \centerline{\epsfxsize=2.5truein \epsfbox{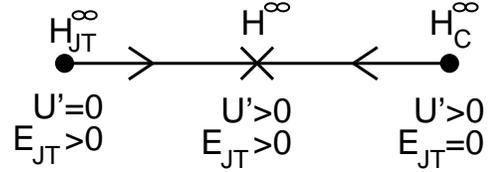} }

  \label{fig1}
  \caption{Schematic representation of the two approaches (Jahn-Teller
    and Coulombic) to the realistic situation based on the 
    simplified model $H^{\infty}$.}
\end{figure}

Another possible simplification could have been obtained by neglecting
the electron-electron interaction in $H$ but keeping the Hund coupling
finite, leading to the following purely JT-phononic model with active
spin degrees of freedom:\cite{yunoki4,hotta5}
\begin{equation}
  H_{\rm JT} = H_{\rm kin} + H_{\rm Hund} + H_{\rm AFM} 
  + H_{\rm el-ph}.
\end{equation}
Note that this model is reduced to $H_{\rm JT}^{\infty}$ when 
the limit $J_{\rm H}$=$\infty$ is taken.
To solve $H_{\rm JT}$, numerical methods such as MC techniques and the
relaxation method have been employed in the past.\cite{yunoki4,hotta5} 
However, here this model will not be discussed explicitly, but the
simplified version $H_{\rm JT}^{\infty}$ will be investigated in the
next section.
Qualitatively, the negligible values of the probability of double
occupancy at large $\lambda$ justifies the neglect of $H_{\rm el-el}$.
In addition, the results of recent studies addressing the $A$-type
AFM-phase of the hole-undoped limit using cooperative JT-phonons have
been shown to be quite similar to those found with pure Coulombic
interactions,\cite{hotta5} the latter treated with the MFA.

Nevertheless, in spite of the above discussed indications that the JT
and Coulomb formalisms lead to similar physics, it would be important
to verify this belief by studying a multi-orbital model with only
Coulombic terms, without the extra approximation of using mean-field
techniques for its analysis.
Of particular relevance is whether PS tendencies and charge ordering
appear in this case, as they do in the JT-phononic model.
This analysis is particularly important since, as explained before,
a mixture of phononic and Coulombic interactions is expected to be
needed for a proper quantitative description of manganites.
For this purpose, yet another simplified model will be analyzed in
this paper:
\begin{equation}
  H_{\rm C} = H_{\rm kin} + H_{\rm el-el}.
\end{equation}
Note that this model cannot be exactly reduced to $H_{\rm C}^{\infty}$,
since the Hund coupling term between $e_{\rm g}$ electrons and 
$t_{\rm 2g}$ spins is not explicitly included. 
The reason for this extra simplification is that the numerical
complexity in the analysis of the model is drastically reduced by
neglecting the localized $t_{\rm 2g}$ spins.
In the FM phase, this is an excellent approximation but not
necessarily for other magnetic arrangements. 
Nevertheless the authors believe that it is important to establish
with accurate numerical techniques
whether the PS tendencies are already present in this simplified
two-orbital models with Coulomb interactions, even if not all degrees
of freedom are incorporated from the outset.
Adding the $S$=3/2 quantum localized spins to the problem would
considerably increase the size of the Hilbert space of the model,
making it intractable with current computational techniques.

\section{Mean-field approximation for $H^{\infty}$}

Even a simplified model such as $H^{\infty}$ is still difficult to
be solved exactly except for some special cases.
Thus, in this section, the MFA is developed for $H^{\infty}$ to
attempt to grasp its essential physics.
Note that even at the mean-field level, due care should be paid to the 
self-consistent treatment for the lift of the double degeneracy 
of $e_g$ electrons.

\subsection{Formulation}

First let us rewrite the electron-phonon term by applying a simple
standard decoupling such as
$q_{2{\bf i}} \tau_{x{\bf i}} \approx 
\langle q_{2{\bf i}} \rangle \tau_{x{\bf i}} + 
q_{2{\bf i}} \langle \tau_{x{\bf i}} \rangle - 
\langle q_{2{\bf i}} \rangle \langle \tau_{x{\bf i}} \rangle$,
where the bracket means the average value using the mean-field
Hamiltonian described below.
By minimizing the phonon energy, the local distortion is determined in
the MFA as $q_{1{\bf i}}$=$-\langle n_{\bf i} \rangle /\beta$,
$q_{2{\bf i}}$=$-\langle \tau_{x{\bf i}} \rangle$, and 
$q_{3{\bf i}}$=$-\langle \tau_{z{\bf i}} \rangle$.
Thus, the electron-phonon term in the MFA is given by
\begin{eqnarray}
 H_{\rm el-ph}^{\rm MF} &=& -2\sum_{\bf i} 
 [ E_{\rm br} \langle n_{{\bf i}} \rangle n_{{\bf i}}
  +E_{\rm JT} (\langle \tau_{x{\bf i}} \rangle \tau_{x{\bf i}}
 +\langle \tau_{z{\bf i}} \rangle \tau_{z{\bf i}})]  \nonumber \\
 &+& \sum_{\bf i}[ E_{\rm br} \langle n_{\bf i} \rangle^2
  + E_{\rm JT} (\langle \tau_{x{\bf i}} \rangle^2
   + \langle \tau_{z{\bf i}} \rangle^2)].
\end{eqnarray}

Now let us turn our attention to the electron-electron interaction
term. 
At a first glance, it appears enough to make a similar decoupling
procedure for $H_{\rm el-el}$.
However, such a decoupling cannot be uniquely carried out, since
$H_{\rm el-el}$ is invariant with respect to 
the choice of $e_g$-electron orbitals
due to the local SU(2) symmetry in the orbital space.
Thus, to obtain the OO state, it is necessary to find the optimal
orbital set by determining the relevant
$e_{\rm g}$-electron orbital self-consistently at each
site. For this purpose, it is useful to express $q_{2{\bf i}}$ and 
$q_{3{\bf i}}$ in polar coordinates as  
\begin{eqnarray}
q_{2{\bf i}} = q_{{\bf i}} \sin \xi_{\bf i},~~
q_{3{\bf i}} = q_{{\bf i}} \cos \xi_{\bf i}.
\end{eqnarray}
In the MFA, $q_{{\bf i}}$ and $\xi_{\bf i}$ can be determined as 
\begin{eqnarray}
  q_{{\bf i}}= \sqrt{\langle \tau_{x{\bf i}} \rangle^2+
    \langle \tau_{z{\bf i}} \rangle^2},
\end{eqnarray}
and 
\begin{eqnarray}
  \xi_{\bf i} = \pi + \tan^{-1} ( \langle \tau_{x{\bf i}} \rangle/
  \langle \tau_{z{\bf i}} \rangle).
\end{eqnarray}
By using the phase $\xi_{\bf i}$,
$c_{{\bf i}{\rm a}}$ and $c_{{\bf i}{\rm b}}$ are transformed into 
${\tilde c}_{{\bf i}{\rm a}}$ and ${\tilde c}_{{\bf i}{\rm b}}$ as 
${\tilde c}_{{\bf i}{\rm a}} = e^{i\xi_{\bf i}/2}
[c_{{\bf i}{\rm a}} \cos (\xi_{\bf i}/2)
+c_{{\bf i}{\rm b}} \sin (\xi_{\bf i}/2)]$
and 
${\tilde c}_{{\bf i}{\rm b}} = e^{i\xi_{\bf i}/2} 
[-c_{{\bf i}{\rm a}} \sin (\xi_{\bf i}/2)+
c_{{\bf i}{\rm b}} \cos (\xi_{\bf i}/2)]$.
Note that the phase factor is needed for the assurance of 
the single-valuedness of the basis function,
leading to the molecular Aharonov-Bohm effect.\cite{hotta}

It should be noted that the phase $\xi_{\bf i}$ determines the
electron orbital set at each site. 
For instance, at $\xi_{\bf i}$=$2\pi/3$, ``a'' and ``b''
denote the $d_{y^2-z^2}$- and $d_{3x^2-r^2}$-orbital, respectively.
In Table.~I, the correspondence between $\xi_{\bf i}$ and the local
orbital is summarized for several important values of $\xi_{\bf i}$.
Note here that $d_{3x^2-r^2}$ and $d_{3y^2-r^2}$ never appear as the
local orbital set. 
In recent publications,\cite{Mutou} those were inadvertently treated
as an orthogonal orbital set to reproduce the experimental results,
but such a treatment is essentially incorrect, since the orbital
ordering is not due to the simple alternation of two arbitrary kinds
of orbitals, as shown in the following.

\medskip
\begin{tabular}{|c|c|c|} \hline
  \makebox[15mm]{$\xi_{\bf i}$} & 
  \makebox[25mm]{a-orbital} & 
  \makebox[25mm]{b-orbital} \\ \hline
  \hline
    $0$    & $x^2-y^2$  & $3z^2-r^2$ \\ \hline 
  $\pi/3$  & $3y^2-r^2$ & $z^2-x^2$  \\ \hline
  $2\pi/3$ & $y^2-z^2$  & $3x^2-r^2$ \\ \hline
  $\pi$    & $3z^2-r^2$ & $x^2-y^2$  \\ \hline 
  $4\pi/3$ & $z^2-x^2$  & $3y^2-r^2$ \\ \hline
  $5\pi/3$ & $3x^2-r^2$ & $y^2-z^2$  \\ \hline
\end{tabular}

\medskip
TABLE I.~Phase $\xi_{\bf i}$ and the corresponding 
$e_{\rm g}$-electron orbitals.
\medskip

By the above transformation, $H_{\rm el-ph}^{\rm MF}$ and 
$H_{\rm el-el}$ are, respectively, rewritten as 
\begin{eqnarray}
 H_{\rm el-ph}^{\rm MF} &=& \sum_{\bf i} 
 \{ E_{\rm br} (-2\langle n_{{\bf i}} \rangle {\tilde n}_{\bf i}
  + \langle n_{\bf i} \rangle^2) \nonumber \\ 
  &+& E_{\rm JT}[2q_{\bf i}
  ({\tilde n}_{{\bf i}{\rm a}}-{\tilde n}_{{\bf i}{\rm b}})
  +q_{\bf i}^2]\},
\end{eqnarray}
and 
\begin{eqnarray}
 H_{\rm el-el}= 
 U'\sum_{\bf i}
 {\tilde n}_{{\bf i}{\rm a}}{\tilde n}_{{\bf i}{\rm b}}
 + V\sum_{\langle {\bf i},{\bf j} \rangle}
 {\tilde n}_{{\bf i}} {\tilde n}_{{\bf j}},
\end{eqnarray}
where ${\tilde n}_{{\bf i}\gamma}
={\tilde c}_{{\bf i} \gamma}^{\dag} {\tilde c}_{{\bf i} \gamma}$ and 
${\tilde n}_{\bf i}
={\tilde n}_{{\bf i}{\rm a}}$+${\tilde n}_{{\bf i}{\rm b}}$.
Note that $H_{\rm el-el}$ is invariant with respect to the choice of 
$\xi_{\bf i}$.
Now let us apply the decoupling procedure as 
${\tilde n}_{{\bf i}{\rm a}} {\tilde n}_{{\bf i}{\rm b}} 
\approx
\langle {\tilde n}_{{\bf i}{\rm a}}\rangle {\tilde n}_{{\bf i}{\rm b}}
+{\tilde n}_{{\bf i}{\rm a}}\langle{\tilde n}_{{\bf i}{\rm b}}\rangle 
-\langle {\tilde n}_{{\bf i}{\rm a}}\rangle
\langle {\tilde n}_{{\bf i}{\rm b}} \rangle$,
by noting the relations 
$\langle {\tilde n}_{{\bf i}{\rm a}} \rangle 
= (\langle n_{{\bf i}} \rangle -q_{\bf i})/2$,
$\langle {\tilde n}_{{\bf i}{\rm b}} \rangle 
= ( \langle n_{{\bf i}} \rangle +q_{\bf i})/2$,
and $n_{\bf i}={\tilde n}_{\bf i}$.
Then, the electron-electron interaction term is given in the MFA as
\begin{eqnarray}
 H_{\rm el-el}^{\rm MF}\! &=&\! (U'/4)\!\sum_{{\bf i}}
 \! [2\langle n_{\bf i} \rangle {\tilde n}_{\bf i}
 \! -\! \langle n_{{\bf i}} \rangle^2
 \!+ \!2q_{\bf i}({\tilde n}_{a{\bf i}}\! - \!{\tilde n}_{b{\bf i}})
 \!+ \! q_{\bf i}^2 ] \nonumber \\ 
 &+& V \sum_{{\bf ia}} [ \langle n_{{\bf i+a}} \rangle 
 {\tilde n}_{{\bf i}} -(1/2) \langle n_{{\bf i+a}} \rangle 
 \langle n_{{\bf i}} \rangle].
\end{eqnarray}
By combining $H_{\rm el-ph}^{\rm MF}$ with $H_{\rm el-el}^{\rm MF}$
and transforming ${\tilde c}_{{\bf i}{\rm a}}$ and 
${\tilde c}_{{\bf i}{\rm b}}$
into the original operators as 
$c_{{\bf i}{\rm a}}$ and $c_{{\bf i}{\rm b}}$,
the mean-field Hamiltonian is finally obtained as 
\begin{eqnarray}
  \label{mfa}
  H^{\infty}_{\rm MF} &=& -\sum_{{\bf ia}\gamma \gamma'}
  t^{\bf a}_{\gamma \gamma'} 
  c_{{\bf i} \gamma}^{\dag} c_{{\bf i+a} \gamma'}
  + J_{\rm AF} \sum_{\langle {\bf i,j} \rangle}
  {\bf S}_{\bf i} \cdot {\bf S}_{\bf j} \nonumber \\ 
  &+& {\tilde E_{\rm JT}} \sum_{\bf i} 
  [-2(\langle \tau_{x{\bf i}} \rangle \tau_{x{\bf i}}
  +\langle \tau_{z{\bf i}} \rangle \tau_{z{\bf i}})
  +\langle \tau_{x{\bf i}} \rangle^2
  + \langle \tau_{z{\bf i}} \rangle^2]    \nonumber \\ 
  &+& \sum_{\bf i} [({\tilde U'}/2) \langle n_{{\bf i}} \rangle
  + V \sum_{\bf a} \langle n_{{\bf i+a}} \rangle]
  n_{{\bf i}}   \nonumber \\
  &-& \sum_{\bf i}[({\tilde U'}/4) \langle n_{{\bf i}} \rangle
  +(V/2)\sum_{\bf a}  \langle n_{{\bf i+a}} \rangle]
  \langle n_{{\bf i}} \rangle,
\end{eqnarray}
where the renormalized JT energy is given by
\begin{eqnarray}
  {\tilde E_{\rm JT}}=E_{\rm JT}+U'/4,
\end{eqnarray}
and the renormalized inter-orbital Coulomb interaction is 
expressed as 
\begin{eqnarray}
  {\tilde U'}=U'- 4E_{\rm br}.
\end{eqnarray}
Physically, the former relation indicates that the JT energy is
effectively enhanced by $U'$.
Namely, the strong on-site Coulombic correlation plays the same role
as that of the JT phonon, at least at the mean-field level, 
indicating that it is not necessary to include $U'$ explicitly in the
models, as has been emphasized by the present authors in several
publications. 
The latter equation for ${\tilde U'}$ means that the one-site 
inter-orbital Coulomb interaction is effectively reduced by the
breathing-mode phonon, since the optical-mode phonon provides an
effective attraction between electrons.
The expected positive value of ${\tilde U'}$ indicates that 
$e_{\rm g}$ electrons dislike double occupancy at the site, since the 
energy loss is proportional to the average local electron number in
the mean-field argument.
Thus, to exploit the gain due to the static JT energy and avoid 
the loss due to the on-site repulsion, 
an $e_{\rm g}$ electron will singly occupy the site.

If $\beta$ is unrealistically small and $E_{\rm br}$ becomes much
larger than $U'$, ${\tilde U'}$ can be negative, leading to an
effective attraction between $e_{\rm g}$ electrons, and
charge-density-wave (CDW) state or $s$-wave superconducting phase 
will appear. 
If the dynamical effect is correctly taken into account beyond the 
adiabatic approximation, the competition among CDW, spin-density-wave,
and $d$-wave superconducting states will occur.\cite{hotta6} 
This is an interesting point, but in the actual manganite $\beta$ is
suggested to be larger than unity.\cite{Iliev}
It has been shown that the breathing-mode distortion is suppressed
for a reasonable choice of parameters, even if it is included in the
calculation.\cite{hotta5}
Thus, in the following, $\beta$ is taken to be infinity for
simplicity, and the breathing mode is dropped from the calculation.

\subsection{JT-phononic vs. Coulombic scenario}

In order to check the validity of the present mean-field treatment,
it is necessary to compare the mean-field result with some 
exact solutions.
For this purpose, numerical techniques are here applied
in small-size clusters. 
For a set of lattice distortions, 
$\{ q_{1{\bf i}}, q_{2{\bf i}}, q_{3{\bf i}} \}$,
the ground-state energy is calculated by using the ED method 
to include exactly the effect of electron correlations.
By searching for the minimum energy, the set of optimal distortions
$\{ q_{1{\bf i}}^{\rm opt}, q_{2{\bf i}}^{\rm opt},
q_{3{\bf i}}^{\rm opt} \}$ is determined.
In this exact treatment, the phase to characterize the local orbital
set is defined by
$\xi_{\bf i}$=
$\tan^{-1}(q_{2{\bf i}}^{\rm opt}/q_{3{\bf i}}^{\rm opt})$.

In this subsection, a small 4-site 1D chain with the periodic boundary
condition is used in order to grasp the most basic aspects of the
problem with a minimum requirement of CPU time. 
The small cluster size and the low dimensionality will be
severe tests for the MFA.
However, if the validity is verified under such severe conditions,
the mean-field result can be more easily accepted in the large cluster
limit and in higher dimensions. 
In order to focus our attention on the roles of JT phonon and
on-site correlations, the $t_{\rm 2g}$ spin configuration is assumed to
be ferromagnetic (thus, $J_{\rm AF}$ will not play an important role) 
and $V$ is set as zero.
The effect of $V$ will be discussed in the next subsection in the
context of the CO state of half-doped manganites.
The realistic hopping matrix set is used for the 1D chain in the
$x$-direction, although the conclusions are independent of the
direction of the 1D chain. 

In Figs.~2(a)-(f), the results for $n$=1 and $E_{\rm JT}$=$t$ are
shown.
Figure 2(a) contains the ground-state energy plotted as a function of 
$U'$ for a fixed value of $E_{\rm JT}$.
In general, the many-body effects cannot be perfectly included within
the MFA. 
However, when $U'$ is introduced to the 1D chain with 
$E_{\rm JT}$$\ne$0, the results becomes exact due to the special
properties of one dimension and the use of the realistic hopping
amplitude.\cite{comment3}
Namely, the hopping direction is restricted only along one axis, and
the orbital set is uniquely determined due to the optimization of the
JT distortion.
In such an orbital polarized situation, the electron correlation
is included exactly even in the mean-field level.
Unfortunately, for 2D or 3D FM phases, there is no guarantee that the 
MFA provides the exact results, as discussed briefly in the final
paragraph, but  the MFA for the Coulomb interactions is still
expected to provide the qualitatively correct tendency due to the
general expectation that the MFA becomes valid in higher dimensions.
However, in the purely JT-phononic model $H_{\rm JT}^{\infty}$,
a remarkable fact appears.
Namely, the MFA is always exact irrespective of the cluster size,
electron density, and dimensionality in $H_{\rm JT}^{\infty}$.
This is quite natural, since the static distortion gives only the
potential energy for the $e_{\rm g}$ electrons, and this fermionic
sector is essentially a one-body problem, even though the potential
should be optimized self-consistently.
The lattice distortion is basically determined by the local electron 
density, and the MFA becomes exact in the static limit for the
distortion. 

\begin{figure}[h]
  \vskip1.7truein
  \hskip-0.2truein
  \centerline{\epsfxsize=4.truein \epsfbox{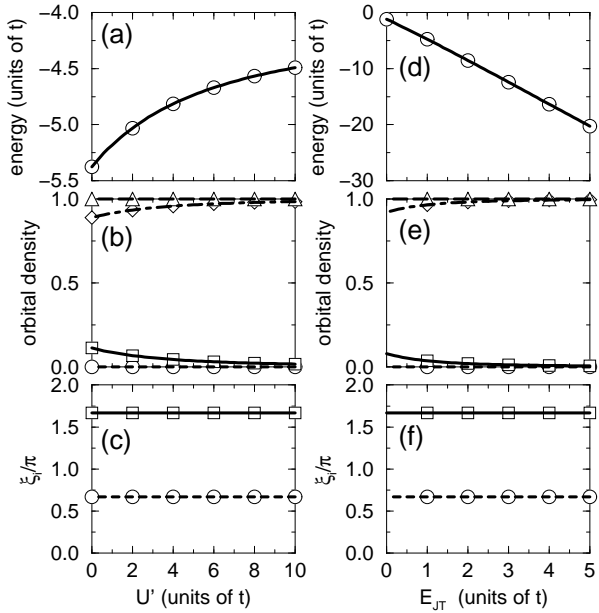} }
  \vskip-1.7truein
  \label{fig2}
\caption{Mean-field and exact-diagonalization results at $n$=1 for
a 4-site 1D chain with the realistic hopping amplitude along the
$x$-axis. 
In all figures, the curves denote the mean-field results and open symbols 
indicate the results obtained by the combination of the exact
diagonalization and the relaxation method in the phonons.
Figures 2(a)-(c) indicate the $U'$-dependence of some quantities for
the special case
$E_{\rm JT}$=$t$, while these same quantites are plotted as a function of 
$E_{\rm JT}$ for $U'$=$5t$ in Figs.~2(d)-(f). 
In (a) and (d), ground-state energies are shown. 
In (b) and (e), orbital densities, 
$\langle {\tilde n}_{{\bf i}{\rm a}} \rangle$ and 
$\langle {\tilde n}_{{\bf i}{\rm b}} \rangle$, are plotted.
The solid curve and open square denote 
$\langle {\tilde n}_{{\bf i}{\rm a}} \rangle$ for site 1.
The broken curve and open circle denote 
$\langle {\tilde n}_{{\bf i}{\rm a}} \rangle$ for site 2.
The dash-broken curve and open diamond indicate 
$\langle {\tilde n}_{{\bf i}{\rm b}} \rangle$ for site 1.
The long-dashed curve and open triangle indicate
$\langle {\tilde n}_{{\bf i}{\rm b}} \rangle$ for site 2.
In (c) and (f), the phases $\xi_{\bf i}$ are plotted.
The solid line and open square denote 
$\xi_{\bf i}$ for site 1.
The broken curve and open circle denote 
$\xi_{\bf i}$ for site 2.
Note that only the results at site 1 and 2 are depicted, since the
results at site 3 and 4 are the same as those at site 1 and 2,
respectively.
}
\end{figure}

For the reasons discussed above the MFA works quite well in the JT
scenario, but this fact does not mean that the obtained state is
trivial, since the orbital degree of freedom is active and several
non-trivial OO phases occur. 
To visualize this result, the orbital densities,
$\langle {\tilde n}_{{\bf i}{\rm a}} \rangle$ and 
$\langle {\tilde n}_{{\bf i}{\rm b}} \rangle$, are shown in Fig.~2(b), 
and the phase $\xi_{\bf i}$ is plotted in Fig.~2(c).
It is observed that $\langle {\tilde n}_{{\bf i}{\rm a}} \rangle$ 
becomes very small and $\langle {\tilde n}_{{\bf i}{\rm b}} \rangle$ 
is almost unity for all the sites.
For even and odd sites, $\xi_{\bf i}$ is given by $2\pi/3$ and
$5\pi/3$, respectively.
Namely, the occupied
b-orbital is $d_{3x^2-r^2}$ at the even sites and $d_{y^2-z^2}$
at the odd sites, respectively.
This is simply the orbital-staggered state in the spin FM phase,
which has been observed in the MC analysis.\cite{yunoki4}
The mechanism of its occurrence is quite simple: 
Imagine the two adjacent sites in which one $e_{\rm g}$ electron is
present at each site.
If the limit $E_{\rm JT}$$\gg$$t$ is considered, the energy gain can be 
evaluated in second order with respect to $t$, 
and this energy is maximized when the occupied orbital in a certain
site is exactly the same as the unoccupied orbital in the neighboring
site. 
Namely, the difference in $\xi_{\bf i}$ between two adjacent sites
should be $\pi$ at $n$=1.
Note that the value of $\xi_{\bf i}$ is determined by the hopping
direction.

Next let us examine the Coulombic scenario, still using a small 4-site 
cluster to compare mean-field results against exact ones.
In Figs.~2(d)-(f), the ground-state energy, orbital densities, 
and the phase are plotted as a function of $E_{\rm JT}$
for a fixed value of $U'$=$5t$. 
Again it can be observed that the MFA works quite well, except for 
the case of $E_{\rm JT}$=0.
In this pure Coulombic limit, the JT distortion does not occur, and
thus, the results are obtained only by the ED for 
$H_{\rm C}^{\infty}$. 
The ground state energy obtained in the MFA converges to the 
exact result in the limit of $E_{\rm JT}$$\rightarrow$ 0.
However, no symbols are shown at $E_{\rm JT}$=0 in Figs.~2(e) and (f),
since the orbital densities and the phase could not be fixed at 
$E_{\rm JT}$=0.
This is due to the fact that the energy is invariant for any choice of 
local orbital, indicating that 
$\langle {\tilde n}_{{\bf i}{\rm a}} \rangle$,
$\langle {\tilde n}_{{\bf i}{\rm b}} \rangle$, and 
$\xi_{\bf i}$ cannot be determined at each site for $E_{\rm JT}$=0.
In other words, the OO state is not uniquely fixed within the
purely Coulombic model in the sense that the a special orbital is not 
specified at each site.
However, the orbital staggered tendency is detected if the orbital
correlation as a function of distance is studied, as will be discussed
in Sec.~VI.

Now the roles of the JT-phonon and on-site correlation are discussed.
At finite electron-phonon coupling, the optimized orbital is
determined, and the MFA provides an essentially exact result for the 
shape of the orbital.
Note that carrying out unbiased MC simulations and using the
relaxation technique are still important tasks, since the present MFA
works quite well only for a $fixed$ $t_{\rm 2g}$ spin background.
Thus, it is an unavoidable step to check whether the assumed 
$t_{\rm 2g}$ spin pattern is really stable or not with the use of 
unbiased techniques by optimizing the lattice distortions as well as
the $t_{\rm 2g}$ spin directions simultaneously.
In this sense, the analytic mean-field approach becomes very powerful
when it is combined with numerical unbiased techniques.

In the purely Coulombic model, however, no special orbital is 
determined at each site due to the local SU(2) symmetry.
Only when the JT distortion is included, the optimal orbital set at
each site is determined from the competition between the kinetic
energy gain and the potential loss due to the lattice distortion.
As a consequence, it appears that the natural starting model to
understand the properties of manganites should be the JT-phononic
model, and it is enough to include the effect of $U'$ on 
$H_{\rm JT}^{\infty}$.
Although the validity of this statement has been shown in a limited
situation in this paper, it is believed to be correct in other cases.
(As for the metal-insulator transition, a comment will be given in the 
final paragraph of this subsection.)
On the other hand, if the starting model is chosen as the purely
Coulombic model, the degeneracy in the orbital space is lifted when
the JT distortion is introduced, indicating that the ground-state
property abruptly changes due to the inclusion of the JT distortion.
However, the on-site correlation is still important to achieve
effectively the strong-coupling region, since $U'$ is renormalized
into the JT energy.

\begin{figure}[h]
  \vskip1.7truein
  \hskip-0.2truein
  \centerline{\epsfxsize=4.truein \epsfbox{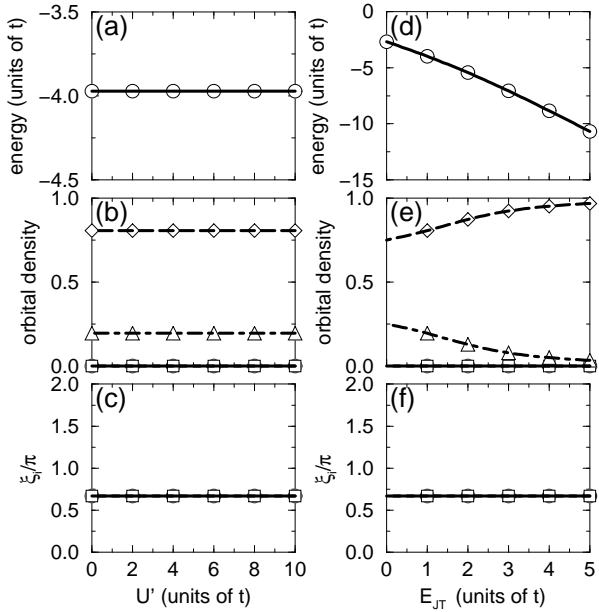} }
  \vskip-1.7truein
  \label{fig3}
  \caption{Mean-field and exact-diagonalization results at $n$=0.5 
    for a 4-site 1D chain with the realistic hopping amplitude along the 
    $x$-axis. The meaning of curves and symbols are the same as those in
    Fig.~2.}
\end{figure}

To check whether the statements in the previous paragraph are 
correct also in the doped case, 
the same analysis is carried out for $n$=0.5, as shown in
Figs.~3(a)-(f).
Note again that the MFA results are always exact and the physical
quantities are continuous as a function of $U'$ for non-zero values of 
$E_{\rm JT}$, but the local orbital set cannot be uniquely determined
for $E_{\rm JT}$=0.
In addition, the results are independent of $U'$ in Figs.~3(a)-(c).
This is understood as follows:
If a unitary transformation is introduced to make the hopping
matrix diagonal, the model $H_{\rm C}^{\infty}$ in the FM-phase is
reduced to
\begin{eqnarray}
  H_{\rm FK} &=& -\sum_{\bf ia}
  (4t/3) a_{{\bf i}}^{\dag} a_{{\bf i+a}}
   + U' \sum_{\bf i} a_{\bf i}^{\dag} a_{\bf i}
   b_{\bf i}^{\dag} b_{\bf i},
\end{eqnarray}
where $a_{\bf i}$ and $b_{\bf i}$ are given by
$a_{\bf i}$=$-(\sqrt{3}/2)c_{{\bf i}a}$+$(1/2)c_{{\bf i}b}$
and 
$b_{\bf i}$=$(1/2)c_{{\bf i}a}$+$(\sqrt{3}/2)c_{{\bf i}b}$,
respectively. 
In this reduced model, only the ``a''-electrons can hop and they 
interact with the localized ``b''-electrons, 
which is just the Falicov-Kimball model.\cite{Falicov}
Of course, this model is obtained trivially if the 1D chain along the 
$z$-direction is considered.
Thus, the ground state at $n$=0.5 is obtained by filling 
the lower ``a''-band up to $n$=0.5 and $U'$ does not work at all.
When the electron-phonon coupling is switched on, the system becomes
insulating,  but the effect of $U'$ is still inactive.
Although the perfect agreement of the mean-field results with the exact
numbers is due to the special properties of the 1D chain,
the JT-phononic model with the realistic hopping appears to contain the
important physics of the problem and the electron correlation is not
as crucially important as naively expected.

As for the orbital densities at $n$=0.5,
$\langle {\tilde n}_{{\bf i}{\rm a}} \rangle$ is always negligible
small, but $\langle {\tilde n}_{{\bf i}{\rm b}} \rangle$ for the even
sites is almost unity, 
while $\langle {\tilde n}_{{\rm b}{\bf i}} \rangle$ for the odd sites
is small, clearly indicating the CO tendency. 
In this case, $\xi_{\bf i}$ is always given by $2\pi/3$ irrespective
of the site position, denoting the occurrence of the ferro ordering of
the $d_{3x^2-r^2}$ orbitals.
This is easily understood by the DE mechanism in the orbital degree of
freedom. 
Namely, the orbital arranges uniformly to improve the kinetic
energy of the $e_{\rm g}$ electrons, just as the spin does in the FM
phase of the doped one-orbital model.

Finally, a brief comment on the MFA in higher dimensions is provided.
For $U'$=0, the 1D system is insulating if $E_{\rm JT}$ is non-zero,
\cite{comment1}
but in higher dimensions, a metal-insulator transition occurs at
some finite value of $E_{\rm JT}$.
If $E_{\rm JT}$ is larger than this critical value, the ground state
wave function will be smoothly changed as $U'$ increases.
However, when $E_{\rm JT}$ is so small that the system is metallic at 
$U'$=0, the inclusion of $U'$ will bring the metal-insulator
transition at some value of $U'$.
The MFA cannot predict correctly this metal-insulator transition
point, but in this paper, the CO state is mainly discussed.
Namely, $U'$ ($E_{\rm JT}$) is included into the insulating phase 
originating from $E_{\rm JT}$($U'$).
Although the properties of the metal-insulator transition in the
manganese oxide are quite interesting, this type of analysis will be
left for the future. 

\section{Charge-Orbital Ordering in half-doped manganites}

In the previous section, the CO-state with a uniform orbital
arrangement has been suggested as the ground-state of the 1D chain 
at $n$=0.5.
However, in the real materials a more complicated situation occurs.
In the narrow-band manganite such as 
La$_{0.5}$Ca$_{0.5}$MnO$_3$ and Nd$_{0.5}$Sr$_{0.5}$MnO$_3$,
the $CE$-type AFM phase is stabilized (See Fig.~4(a)).
In this structure, the $t_{\rm 2g}$ spins form a ferromagnetic array
along a zigzag 1D path, the CO state appears with the checkerboard
pattern, and the $(3x^2-r^2)$/$(3y^2-r^2)$-type orbital ordering is
concomitant to this CO state.
The origin of this complex spin-charge-orbital structure
has been recently clarified on the basis of the topology of the 
zigzag structure.\cite{hotta2}
Along the $z$-axis, the $CE$-pattern stacks with the same CO/OO
structure, but the coupling of $t_{\rm 2g}$ spins along the $z$-axis
is antiferromagnetic. 
If only the bi-partite charge structure in the $x$-$y$ plane is
considered, naively the nearest-neighbor Coulomb interaction $V$ 
seems to be the key issue in the formation of the CO/OO
phase.\cite{Mutou}
However, if $V$ is strong enough to bring the CO structure in the
$x$-$y$ plane, the 3D Wigner-crystal structure should appear in the
cubic lattice, but this is not observed in the real system.
Thus, $V$ is not the only ingredient needed for the stabilization of
the CO structure of half-doped manganites, as already discussed in
the Introduction. 
What other terms in the Hamiltonian may create a charge-stacked phase?

\begin{figure}[h]
  \vskip1.8truein
  \hskip-0.2truein
  \centerline{\epsfxsize=4.truein \epsfbox{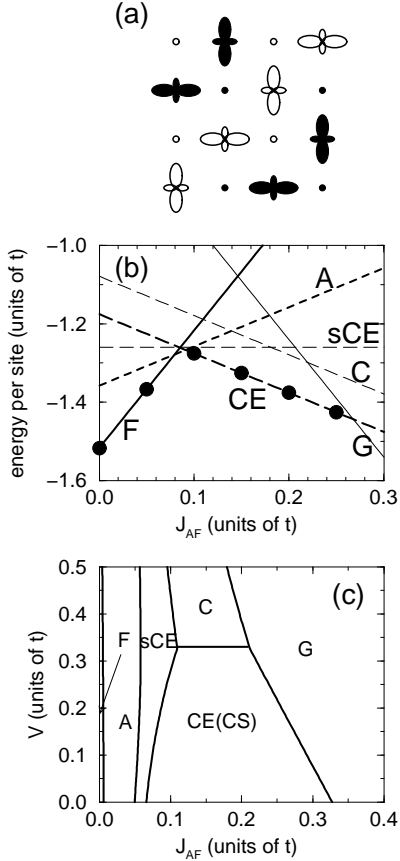} }
  \vskip-.5truein
  \label{fig4}
  \caption{
    (a) Schematic representation of the 
    spin, charge, and orbital ordering for the 
    $CE$-type AFM structure. Solid and open symbols indicate up and down 
    $t_{\rm 2g}$ spins, respectively.
    The lobes indicates  $(3x^2-r^2)$- or $(3y^2-r^2)$-orbitals at
    the Mn$^{3+}$ sites, while the circles denote Mn$^{4+}$ sites.
    (b) Energy per site as a function of $J_{\rm AF}$ for $\lambda$=1.6
    and  $U'$=$0$. 
    The curves denote the mean-field results and the solid symbols
    indicate the energy obtained by the relaxation method.
    Thick solid, thick broken, thin broken, thick dashed, thin dashed,
    thin broken, and thin solid lines denotes FM, $A$-type, shifted
    $CE$-type, $CE$-type, $C$-type, and $G$-type states, respectively.
    (c) Phase diagram in the $(J_{\rm AF},V)$ plane. Note that the
    charge-stacked structure along the $z$-axis can be observed 
    only in the $CE$-type AFM phase.}
\end{figure}

In order to clarify this point, in this subsection 4$\times$4$\times$4 
lattices with the periodic boundary condition are studied as a simple
representation of half-doped manganites on the basis of $H^{\infty}$.
In Fig.~4(b), energies for several magnetic phases are plotted 
as a function of $J_{\rm AF}$ for $\lambda=1.6$.\cite{hotta4}
At this stage in the calculations, $V$ is set to zero and its effect
will be discussed later.
The results of Fig.~4(b) are obtained from the JT-phononic model, but 
as mentioned in the previous section, it can be interpreted that the
effect of $U'$ is included effectively in the electron-phonon
coupling.  
Although the effect of $U'$ also appears through the non-trivial term
$({\tilde U'}/2)\sum_{\bf i} \langle n_{\bf i} \rangle n_{\bf i}$, 
this term essentially indicates the prohibition of double occupancy,
and does not lead to qualitative changes in the results obtained from
the JT-phononic model, as long as $\beta$ is larger than unity and
${\tilde U'}$ is positive. 
Thus, the results below can be interpreted as arising from a purely JT
calculation or a mixture of JT and Coulomb interactions within a
mean-field technique.
The curves shown are the results in the MFA obtained for several fixed
$t_{\rm 2g}$ spin  patterns, while the solid circles are obtained by
the optimization of both the local JT distortion and the $t_{\rm 2g}$ 
spin angles simultaneously.
The agreement between the analytic and numerical results is excellent,
indicating that the MFA works quite well in the JT-phononic model.
For $J_{\rm AF}$$\alt$0.1$t$, a metallic 3D FM-phase 
is stabilized, since this spin pattern optimizes 
the kinetic energy.
In a very narrow region around $J_{\rm AF}$$\approx$0.1$t$, the
$A$-type AFM phase occurs.
This phase is metallic in the present intermediate coupling, but 
the uniform $(x^2-y^2)$-type orbital ordering appears.
In the wider region 0.1$t$$\alt$$J_{\rm AF}$$\alt$0.25$t$,
the $CE$-type AFM structure is the ground state.
For unrealistic large values of $J_{\rm AF}$, 
the $G$-type AFM phase is stabilized to gain the magnetic energy.

Now let us focus our attention on the stabilization of the $CE$-type
phase. 
Due to the competition between the kinetic energy of the $e_{\rm g}$
electrons and the magnetic energy of the $t_{\rm 2g}$ spins,
the one-dimensional stripe-like AFM configuration occurs,\cite{hotta2}
in which arrays of $t_{\rm 2g}$ spins order ferromagnetically along
some particular 1D paths. 
Precisely the shape of the 1D FM path, the zigzag path, is a key to
understand the $CE$-type phase.
To see this, let us compare the $CE$-state with the $C$-type AFM
state, which is characterized by the straight-line FM path.
Although the energy of the $C$-type structure has the same slope 
as a function of $J_{\rm AF}$ as
the $CE$-type, it is not the ground-state, as shown in Fig.~4(b).
As for the orbital arrangement,
the $(3x^2-r^2)$-type OO occurs in the $C$-type, while 
the $(3x^2-r^2)/(3y^2-r^2)$-type orbital arrangement appears
in the $CE$-type state.
At a first glance, these two OO-states seem to be quite different,
but if the concept of the orbital DE mechanism is employed,
it is reasonable that there is no essential distinction between them.
Namely, in both cases, the $e_{\rm g}$ electron orbital is always
polarized along the hopping direction of the 1D paths.

An essential point for the stabilization of the charge-stacked
$CE$-type AFM phase is the difference of the $topology$ of the 1D paths
between the straight and zigzag shapes.
Mathematically, the topology of the 1D path can be characterized by 
``the winding number'' $w$\cite{hotta} defined from the Berry phase
connection of the $e_{\rm g}$-electron wave function along the hopping
path.\cite{koizumi} 
As shown in Ref.~\onlinecite{hotta2},
$w$ is decomposed into two terms as $w$=$w_{\rm g}$+$w_{\rm t}$.
The former, $w_{\rm g}$, is called the geometric term,\cite{takada}
which is $0$ ($1$) corresponding to the ferro- (antiferro-)
arrangement in the orbital sector along the 1D path.
As discussed above, due to the orbital DE mechanism, $w_{\rm g}$ is
zero in the doped case irrespective of the shape of the hopping path.
The difference between the straight and zigzag paths appears in
$w_{\rm t}$, expressed as $w_{\rm t}$=$N_{\rm v}/2$, where 
$N_{\rm v}$ is the number of vertices appearing in the unit 
of the 1D chain.\cite{hotta2}
Since $w_{\rm t}$ is determined only by the shape of the 1D path,
it is called the topological term.
Of course, $N_{\rm v}$ is zero for the straight path, while 
$N_{\rm v}$ is equal to 2 for the zigzag path. 
Thus, it is obtained that $w$=0 and $1$ are the values 
for the $C$- and $CE$-type AFM
phases, respectively.

To understand why the zigzag chain with $w$=1 has the lower energy,
it is useful to consider the situation $U'$=$E_{\rm JT}$=0, 
in which the $CE$-type AFM spin structure characterized by the zigzag
1D chain is stabilized in the picture of a band
insulator.\cite{hotta2,terakura,brink}
Due to the periodic change along the zigzag path in the hopping
amplitude as 
\{ $\cdots$, $t_{\gamma\gamma'}^{\bf x}$, $t_{\gamma\gamma'}^{\bf x}$,
$t_{\gamma\gamma'}^{\bf y}$, $t_{\gamma\gamma'}^{\bf y}$, $\cdots$ \},
the system becomes the band insulator. 
Although the difference in the hopping amplitudes in the $x$- and
$y$-directions is only a phase factor in 
$t_{{\rm ab}}^{\bf y}$ and $t_{{\rm ab}}^{\bf x}$,
a bandgap of the order of $t$ develops.
On the other hand, note that the $C$-type AFM states characterized by
the straight path is metallic, since there is no change in
the hopping amplitude from bond to bond.
In Ref.~\onlinecite{hotta2}, it has been clearly shown that the
$CE$-type structure has the largest bandgap among all the possible
types of zigzag hopping paths at $n$=0.5.
Moreover, in Ref.~\onlinecite{hotta2}, if the JT energy is switched on
in this phase, it has been shown that the $CE$-type phase continues to
be the ground state and charge-ordering appears with the 
$(3x^2-r^2)/(3y^2-r^2)$-type orbital-ordering. 
Thus, based on the band-insulator picture and the spirit of the 
adiabatic continuation, the $CE$-type AFM structure characterized
by the 1D zigzag FM chain is the ground-state.

Now the physical meaning of the energy difference between $CE$- and
$C$-type states is discussed using the the concept of the winding
number.  
For this purpose, an analogy with a typical spin problem is quite
useful, as suggested by Takada {\it et al.}\cite{takada}
In the spin problem, by classifying the states with the total spin $S$ 
which is a conserved quantity, the exchange energy is defined by the
difference between the energies of the singlet ($S$=0) 
and triplet ($S$=1) states.
In the present problem, $w$ is the topological entity and the
conserved quantity.
Thus, the energy difference between the states with $w=1$ and $w=0$,
$J_{\rm w}$, is expected to play a similar role as the exchange energy
in the spin problem.
As for the magnitude of $J_{\rm w}$, it can be of the order of $0.1t$
from the analysis of the two-site problem,\cite{takada}
although it depends on the value of $E_{\rm JT}$.

Another phase competing with the $CE$-type state is the shifted
$CE$-type (s$CE$) structure.\cite{hotta3,hotta4}
This is also obtained by the stacking of Fig.~4(a) along the $z$-axis,
but one-lattice spacing shifted in the $x$- (or $y$-)direction.
Due to this shift, the number of FM and AFM bonds becomes equal, and
the magnetic energy is exactly canceled.
In fact, the energy for the s$CE$ structure is independent of $J_{\rm AF}$.
The $CE$-type phase is stabilized against the s$CE$-structure by the 
magnetic energy, as observed in Fig.~4(b), showing the key role played
by $J_{AF}$ in models for manganites and likely in the real compounds 
as well.

Let us consider now the effect of $V$ on the CO states. 
For this purpose, the mean-field calculation is carried out based on 
$H_{\rm MF}^{\infty}$ for a reasonable parameter set \cite{comment2}
such as $t$=0.5eV,\cite{hopping} $E_{\rm JT}$=0.25eV,\cite{Shen} and
$U'$=5eV.\cite{ishihara}
The phase diagram in the $(J_{\rm AF},V)$ plane is shown 
in Fig.~4(c). From this figure it is clear that the CS structure
occurs for the $CE$-type AFM phase in a broad region of parameter
space, as deduced from the Fourier transform of the charge correlation
which has an observed peak at $(\pi,\pi,0)$ in the $CE$-type state,
while a peak appears at $(\pi,\pi,\pi)$ in the s$CE$ and $C$-type AFM
states.\cite{hotta3}
A remarkable fact of Fig.~4(c) is that the CS structure is robust
against the inclusion of $V$.
The boundary between the s$CE$- and $CE$-phases is qualitatively
understood by the balance of the magnetic energy gain and the charge
repulsion loss.\cite{hotta4}
On the other hand, the phase boundary between the $CE$- and $C$-type
AFM states is independent of $J_{\rm AF}$, since those two states
have the same magnetic energy.
As mentioned above, in this case, the energy $J_{\rm w}$ 
due to the difference in the topology of 1D path stabilizes the CS
structure in spite of the charge repulsion loss.
In fact, the phase boundary exists around $V$$\approx$0.3$t$, which is
the same order as $J_{\rm w}$.
Although it is difficult to know the exact value of $V$ in the actual
material, if the simple screened Coulomb interaction is estimated
using the large dielectric constant of manganites,\cite{arima} 
$V/t$ is estimated to be 0.1$\sim$0.2,\cite{yunoki4} i.e., 
inside the CS region in our phase diagram.

Finally, a comment on the stabilization of the $CE$-type structure 
in the purely Coulombic model is provided.
As discussed above, in the case of $U'$=$E_{\rm JT}$=0, the $CE$-type 
AFM spin structure is found to be the ground-state in the
band-insulator picture. 
If $U'$ is smoothly switched-on in this phase, still keeping
$E_{\rm JT}$=0, the $CE$-type phase
still continues to be the ground state and charge-ordering appears
even without the help of $V$, since the local charge in the straight
segment of the zigzag chain is larger than that at the corner
site.\cite{brink}
The ground state energy for the zigzag chain is lower than that for
the straight chain, and its difference is again of the order of 0.1$t$.
Since this energy difference can compensate the energy loss due to $V$
in the CS structure, it would be possible to understand the CS
structure even in the purely Coulombic model on the same topological
argument as carried out for the JT-phononic model.
Thus, it is possible to fill the 3D cubic lattice by the zigzag 1D chains
stacked in the $b$- and $c$-axis directions,
with the same charge ordering but antiparallel $t_{2g}$-spin
directions across those 1D chains.
Note, however, that the energy is invariant for the choice of
local $e_{\rm g}$-electron orbital and the
$(3x^2-r^2)/(3y^2-r^2)$-type orbital arrangement $cannot$ be specified 
in the purely Coulombic model.
On the other hand, in the pure JT-phononic model, the $CE$-type AFM
spin structure, the charge-stacked CO state, and the
$(3x^2-r^2)/(3y^2-r^2)$-type OO-phase have been fully understood,
as explained before in this section. 
Thus, based on these results, these authors believe that the purely
JT-phononic model is more effective than the purely Coulombic model
for the theoretical investigation of manganites, although certainly
both lead to very similar physics.

\section{Phase-Separation Tendency}

In previous investigations using the unbiased MC simulations,
the PS tendency has been clearly established in manganite models.
\cite{yunoki,yunoki2,yunoki3,yunoki4}
In the simulations, PS appeared both in the one-orbital FM Kondo
model\cite{yunoki,yunoki2,yunoki3} and the two-orbital JT-phononic
models.\cite{yunoki4}
PS occurs due to the balance between the kinetic energy 
of the $e_{\rm g}$ electrons and the potential energy due to the
background, either the $t_{\rm 2g}$ spins or the JT distortion.
Due to this difference in the origin of the background potential,
two-types of PS exist, spin driven and orbital driven.
In the following, the 1D case is used for simplicity.
In higher dimensions, the situation will be more complicated,
but the essential physics is expected to be captured in the 1D case.

One type of PS appears in the two-orbital model driven by the spin
sector between the spin FM and AFM phases, mainly in the region
between $x$$\approx$0.5 and $x$=1.0.
To improve the kinetic energy, in this doped situation the 
$t_{\rm 2g}$ spins and $e_{\rm g}$-electron orbital tend to array in a 
ferro-manner, but in the heavily doped region close to $x$=1.0, the
$t_{\rm 2g}$ spins order antiferromagnetically to gain the magnetic
energy which dominates over the kinetic energy.
Thus, in this case, the PS between FM/OF- and AFM/OF-states appears,
as clearly indicated in the previous MC simulations.
Note that the acronym in front of the slash indicates the spin state, 
while the acronym after the slash denotes the orbital arrangement,
e.g., FM/OF indicates the spin FM and orbital ferro state.
Note that this PS is possible in the one-orbital model as well, since 
the orbital degrees of freedom is not active for this case.

Another variety of PS is related to the orbital degrees of freedom,
which is tightly coupled to the JT distortion, which mainly exists
between $x$=0 and $x$$\approx$0.5 in the two-orbitals model.  
In the undoped situation and for reasonable values of $J_{\rm AF}$,
the FM/OAF-state is the ground state, as shown in the previous
section.
Also in the unbiased MC simulation, this phase has been
obtained.\cite{yunoki4}
When holes are doped, the spin structure is still FM, but the orbital
arrangement becomes uniform to gain the kinetic energy by polarizing
the orbitals along the hopping direction (orbital DE mechanism).
Thus, in this case, the PS between FM/OAF- and FM/OF-states appears,
as suggested in the previous works for the two-orbital
model.\cite{yunoki4}

In the MC simulation, the grand canonical ensemble has been used and
the chemical potential $\mu$ was tuned to obtain the target electron
density.
In such a calculation, the presence of PS was clearly observed by
monitoring the density vs. $\mu$.
There is a range of densities that cannot be stabilized, no matter how 
carefully $\mu$ is fine-tuned.
In other words, the plot $\langle n \rangle$ as a function of $\mu$
presents a discontinuity at some particular value of $\mu$.
This effect occurs for the one- and two-orbital models, the latter
with JT-phonons, and in all dimensions of interest.
Further evidence of mixed-phase tendencies can be obtained by
monitoring the density as a function of MC-time. 
In a stable regime, the density does not change much between MC
configurations, but in the PS region the fluctuations are very strong.

\begin{figure}[h]
  \vskip1.8truein
  \hskip-0.2truein
  \centerline{\epsfxsize=4.truein \epsfbox{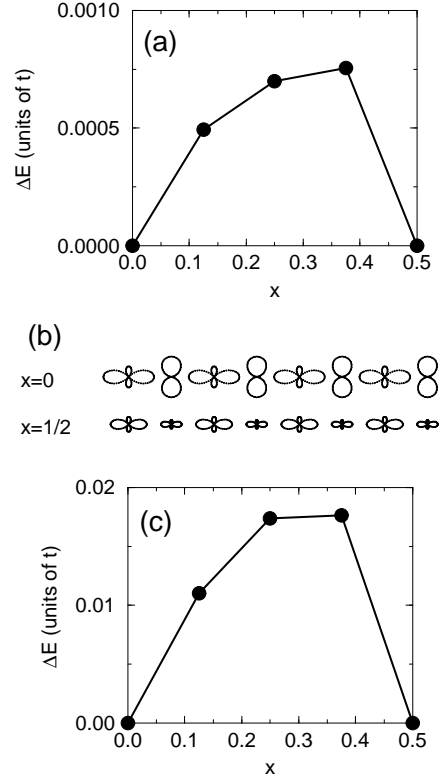} }
  \vskip-.8truein
  \label{fig5}
  \caption{(a) Energy difference $\Delta E$ as a function of $x$ for
    a 16-site 1D chain with $U'$=0 and $E_{\rm JT}$=1.2.
    The definition of $\Delta E$ is shown in the text.
    (b) Orbital density $\langle {\tilde n}_{{\bf i}{\rm b}} \rangle$ for 
    $x$=0 (upper chain) and $0.5$ (lower chain).
    Note that the b-orbital in the $x$-$y$ plane is depicted and its 
    size is proportional to 
    $\langle {\tilde n}_{{\bf i}{\rm b}} \rangle$.
    (c) Energy difference $\Delta E$ as a function of $x$ for
    a 16-site 1D chain with $U'$=10 and $E_{\rm JT}$=0.}
\end{figure}

In the present MFA, the electron number $n$ has been fixed.
For a given $n$, the optimized structure both for the $t_{\rm 2g}$
spins and the lattice distortion are determined by the analytic MFA
and the numerical technique.
To understand the PS tendency in this formalism, it is necessary to
check the stability of the obtained phase by calculating the 
ground-state energy $E_0$ as a function of $n$.
Namely, if a negative curvature is obtained, i.e., 
$\partial ^2 E_0/\partial n^2$$<$0, such a phase is unstable even if
it is the ground state at a fixed $n$.
If the PS tendency is included in the present model, the FM/OF states
between $x$=0 and $x$=0.5 should be unstable and the mixed phase
should have lower energy than the states obtained by the MFA, in order 
to reproduce the MC results.
In order to verify this, the energy difference $\Delta E(x)$ is
plotted as a function of $x$ for $E_{\rm JT}$=1.2 in a 16-site spin-FM
1D-chain with the realistic hopping along the $x$-axis.
Here $\Delta E(x)$=$E_0(x)$$-$$\epsilon x$ with 
$\epsilon$=$2[E_0(x$=$0)$$-$$E_0(x$=$0.5)]$.
If $\Delta E(x)$ is positive, the homogeneous state is unstable and,
instead, the mixed-phase appears. 
Note that the state at $x$=0.5 is stable.
The result is shown in Fig.~5(a), in which positive $\Delta E(x)$ is
indeed observed between $x$=0 and $x$=0.5.
In Fig.~5(b), the orbital arrangements at $x$=0 and $x$=0.5 are shown
by depicting the shape of the occupied b-orbital in the $x$-$y$
plane.
It should be noted that the size of the orbital is proportional to
$\langle {\tilde n}_{b{\bf i}} \rangle$.
This result agrees very well with the previous MC calculation, and the
PS tendency is believed to be definitely established in the
JT-phononic model.

Now let us analyze whether it is possible to detect the PS tendency in the
purely Coulombic model or not.
For this purpose, $\Delta E(x)$ is evaluated using a 16-site spin-FM
1D-chain with the realistic hopping along the $x$-axis for 
$E_{\rm JT}=0$ and $U'=10t$ by using the mean-field Hamiltonian
Eq.~(\ref{mfa}).
As shown in Fig.~5(c), again the positive $\Delta E$ for $x$ between 
0 and 0.5 can be observed, indicating clearly the PS tendency.
If the present mean-field Hamiltonian is accepted, this result is 
quite natural, since $U'$ is included effectively in the coupling
between the $e_{\rm g}$ electrons and JT distortion. 
However, one may consider that this is just an artifact due to the MFA.
Thus, in the following section, it is explicitly shown that this PS
tendency in the purely Coulombic model is not due to particular 
properties of the MFA by performing the DMRG calculation 
in the 1D Hubbard-like model. 

In short, it is quite interesting to observe that the MFA can properly
reproduce the PS tendencies found using other more sophisticated 
techniques. Although the calculation is fairly simple and handy
the obtained result is physically meaningful and reliable.
More investigations on the PS tendency by using the MFA in higher
dimensions would be certainly interesting since the MC simulations
become increasingly difficult as such dimension grows, but this point 
will be discussed in future publications.

\section{DMRG result for $H_{\rm C}$}

The purpose of this section is to continue the analysis of 
the reduced Hamiltonians of Section II, this time focusing on
the purely Coulombic model $H_{\rm C}$. 
The multi-orbital Hubbard model has been addressed before using a
variety of approximations,\cite{rts}
but here care must be taken to select the appropriate technique
accurate enough to search for physics similar to the results 
obtained with JT-phonons. 
The use of unbiased methods is particularly important for subtle
issues such as PS.
For this purpose, the model $H_{\rm C}$ described in Section II has
been studied here with the DMRG method,\cite{white} supplemented by
the ED technique, keeping the truncation errors around $10^{-6}$ by
using typically 120 states on intermediate size chains with open
boundary conditions.
The restriction to work in the 1D system is not severe in view of the
results of previous sections and those of
Refs.\onlinecite{yunoki,yunoki2,yunoki3},
that showed a strong similarity between one, two and three dimensions, 
at least regarding the rough features of the ground state.
Actually, it is important to remark that the phase diagrams described
in previous work by our group are mostly based on evidence coming from
the short-distance correlations calculated in our study which are not 
expected to depend strongly on the dimensionality.
In 1D systems, the existence of genuine long-range order vs. slow
power-law decay of correlations is a subtle issue beyond the goals of 
the present analysis.

In this context, the static observables studied with the DMRG method
are the spin structure factor 
\begin{equation}
  S(k) = \frac{1}{L} \sum_{j,m}
\langle {\bf s}_{j} \cdot {\bf s}_{m} \rangle \: e^{i(j-m)k},
\end{equation}
the charge structure factor
\begin{equation}
  N(k) = \frac{1}{L} \sum_{j,m}
  \langle \rho_{j}\: \rho_{m} \rangle \: e^{i(j-m)k},
\end{equation}
and the orbital structure factor
\begin{equation}
  T^{z}(k) = \frac{1}{L} \sum_{j,m}
  \langle T^{z}_{j}\: T^{z}_{m} \rangle \: e^{i(j-m)k},
\end{equation}
where $k$ is momentum, $j$ and $m$ denote site positions,
$L$ is the length of the 1D chain,
$T^{z}_{j}$=$(\rho_{j{\rm a}}$$-$$\rho_{j{\rm b}})/2$,
and $\rho_{j\gamma}$=$\sum_{\sigma}\rho_{j\gamma\sigma}$.
It is important to remark that the analysis described below has
focussed on the analogies with the results found using JT-phonons, 
especially regarding PS and the existence and properties of the
spin-AF orbital-staggered phase at density $n$=1.
In the several other interesting phases reported in this section, our
effort has been limited to the description of their main features
regarding their spin, charge and orbital characteristics, postponing 
for a future publication a more detailed analysis of its origin and
possible relevance to experiments. 
In order to focus on the similarity in the effects of the on-site
correlation and the JT phonons, $V$ is set to be zero for the time
being. 
The influence of $V$ will be discussed later.
Note that the energy unit is $t$ in this section, as in the previous
ones.

\subsection{Unit Hopping Matrix}

Thus far the realistic hopping matrix with non-zero off-diagonal
hopping amplitude has been used.
This hopping matrix is quite important to understand the properties
of manganites, but the analysis is sometimes complicated due to the
non-zero off-diagonal element.
In order to analyze the two-orbital model more easily,
it is useful to introduce first a simple unit hopping matrix for any
direction, given by
\begin{eqnarray}
  t_{\gamma \gamma'}^{\bf a}=t \delta_{\gamma \gamma'},
\end{eqnarray}
where $\delta_{ij}$ is the Kronecker's delta.
Besides its simplicity, this hopping amplitude has been used before in 
the search for the FM state in the multi-orbital Hubbard model with a
variety of techniques,\cite{rts} 
and thus, it is meaningful to investigate its properties. 
In addition, it will be theoretically interesting to compare results
using different hopping sets, to study the dependence of the
ground-state properties with those amplitudes. It will actually be
concluded that the use of realistic hoppings is very important
in this context.

\subsubsection{Case of $n$=1}

Consider first the special case of density $n$=1.
Using the DMRG and ED methods a large set of couplings $(U',J)$
have been investigated, but here only the most representative results
are presented as examples.
In Fig.~6(a), the spin structure factor $S(k)$ is shown at fixed 
$U'$=$10$, for particular values of $J$, and three regimes are clearly 
identified.
At small $J$ compared with $U'$, the spin sector has incommensurate
characteristics, with a maximum at momentum $k$=$\pi/2$. 
As $J$ grows, an abrupt transition to the FM state is observed, with
$S(k)$ now peaked at zero momentum.
This last regime occurs in a robust window of $J$.
At $J$$\sim$15, the spin sector becomes incommensurate again with a
peak in $S(k)$ at $k$=$\pi/2$.
The charge structure factor $N(k)$ is shown in Fig.~6(b) for the same
set of parameters. 
It is observed that at $J$=2 and 5 the charge correlations are not
enhanced since only a broad peak with low-intensity appears at
$k$=$\pi$.
However, for $J$=12 and 20 indications of charge-ordering tendencies
are found, since now the $k$=$\pi$ result is clearly enhanced,
indicating a structure with an alternation between the even- and
odd-sites of the chain. 
Note that $J$=5 and 12 have different characteristics when analyzed
using $N(k)$, while they are both ferromagnetic according to $S(k)$.

In Fig.~7(a), the orbital structure factor $T^z(k)$ is presented for
the same set of couplings as used in Figs.~6(a) and (b). 
In this case, enhanced orbital correlations exist in the ground state
at $J$=2 due to the robust values that $T^z(k)$ has, particularly at
$k$=$\pi$. 
This is indicative of a $staggered$ arrangement of orbitals, similar
to the results discussed before in Sec.~III.
A qualitatively similar but even more prominent effect occurs at $J$=5. 
At $J$=12 and 20, $T^z(\pi)$ is relatively small and there is no
noticeable indication of enhanced orbital correlations. 
The transition from the low-$J$ regime, exemplified by $J$=2, to the 
intermediate one ($J$=5) is abrupt, as already observed in the study 
of the spin structure factor.

\ 
\begin{figure}[h]
  \vskip-0.8truein
  \centerline{\epsfxsize=3truein \epsfbox{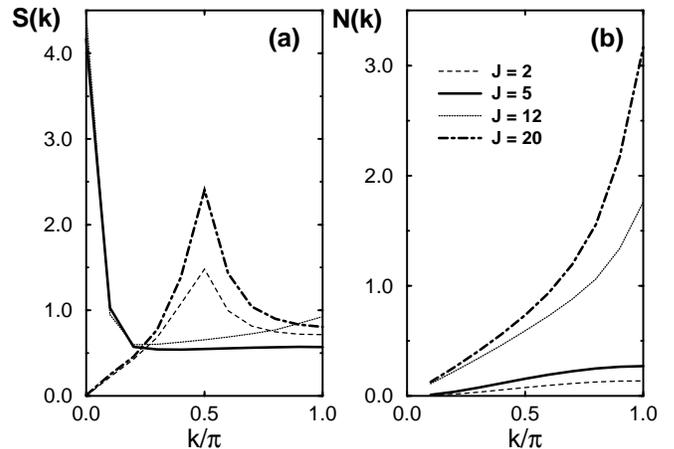} }
  \vskip-0.7truein
  \label{fig6}
  \caption{DMRG results for (a) spin structure factor $S(k)$ and (b)
    charge structure factor $N(k)$ as a function of $k$ for $H_{\rm C}$
    with the unit hopping matrix in the 20-site 1D chain for $U'$=10.
    The dashed, solid, dotted, and dotted-dashed lines correspond to 
    $J$=2, 5, 12 and 20, respectively. 
    }
\end{figure}

\begin{figure}[h]
  \vskip-0.8truein
  \centerline{\epsfxsize=3truein \epsfbox{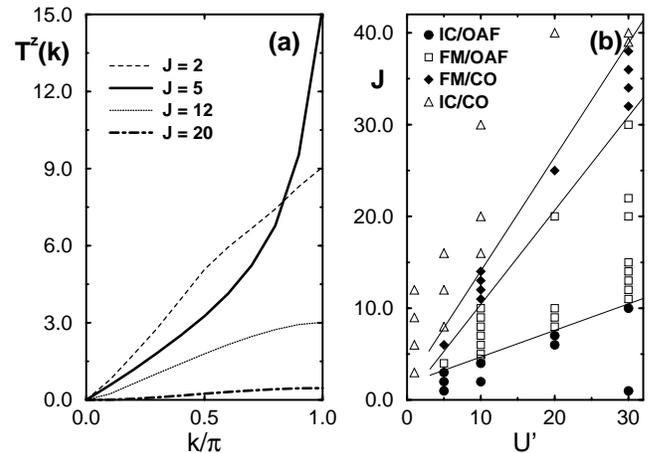} }
  \vskip-0.7truein
  \label{fig7}
  \caption{(a) Orbital structure factor $T^z(k)$ vs. $k$ using the same 
    parameters as in Fig.~6.
    (b) Phase diagram of $H_{\rm C}$ for the unit-matrix hopping amplitude 
    at $n$=1.
    The four phases are IC/OAF, FM/OAF, FM/CO, and IC/CO.
    Here ``IC'' means spin-incommensurate, 
    ``FM'' means spin-ferromagnetic, 
    ``OAF'' denotes orbital-staggered,
    and ``CO'' indicates charge-ordred.}
\end{figure}

Repeating a similar analysis for several other values of $U'$ and J
allowed us to sketch a phase diagram for the two-orbital Hubbard model
at density $n$=1, which is shown in Fig.~7(b). 
Since the complexity of the Hamiltonian does not allow us to perform a
careful finite-size study to determine the dominant correlations in
the bulk ground-state, the phase diagram of Fig.~7(b) should be
considered only qualitative. Nevertheless, since for cluster sizes
smaller than $L$=20 the results were found to be similar, no large
size-effects are anticipated. 
The four identified regions are marked.
Note that the emphasis has been given to the determination of the
boundary of the spin-ferromagnetic orbital-ordered phase due to its
similarities with the analogous phase reported in the MC studies of the
two-orbital model with JT phonons,\cite{yunoki4}
and in the MFA of the previous section. 
The boundaries of the other regimes are less accurately determined,
particularly at small values of $J$ and $U'$.

Let us consider the origin of the many phases observed in Fig.~7.
First address the CO regime which appears above the line $J$=$U'$.
The intuitive explanation for this behavior is simple.
The inter-orbital exchange interaction $J$ actually provides an
attractive interaction to the electrons, which try to doubly populate
half the sites of the chains when this interaction dominates.
On the other hand, the inter-orbital repulsion $U'$ prevents double
occupancy of two orbitals at the same site. 
Then, a competition between the two occurs naturally. 
If $J$$>$$U'$, which is an unphysical limit, then charge-ordering occurs
with roughly half the sites having two particles and the other half
none.
To allow for some nonzero electronic kinetic energy, the doubly
occupied sites do not cluster together but are spread on the chain.
The spin and charge pattern compatible with the results of Figs.6 and
7 in the CO regime has a unit cell of size four lattice spacings,
and an arrangement (2u,0,2d,0) where ``2u'' denotes two spins up,
``0'' is an empty site, and ``2d'' are two spins down. 
Spin staggered patterns appear frequently in the CO states.

The region $U'$$>$$J$ is physically more interesting, and connected
with the results observed for the model with JT-phonons at the same
density.  
The spin FM characteristic of the FM/OAF phase is believed to be
caused by the optimization of the kinetic energy by spin alignment.
The influence of the attractive coupling $J$ is also very important for 
the stabilization of this phase, since it also favors spin alignment
at every site. 
The orbital-staggered pattern allows for some mobility of the
electrons from site to site, while an orbital-uniform arrangement
would not allow for that movement at density $n$=1, if all spins were
aligned and when the unit-matrix hopping is used. 
The charge is spread uniformly, i.e., no charge-ordering exists in
this phase. 
However, below but close to the line $J$=$U'$, charge correlations are 
enhanced due to the proximity to the CO-regime.
In this subregion, the orbital correlations are suppressed compared
with the results observed at the lower $J$ end of the FM/OAF phase,
where they are maximized. 
Overall, it is clear that the FM/OAF state has characteristics very
$similar$ to those observed when JT-phonons are used to mediate the
interaction between electrons (See Sec.~III).
Then, this phase appears prominently both in studies with phonons and
Coulombic interactions, and likely it will be stable when a mixture of
the two terms is used.

As $J$ is decreased for fixed $U'$, the ferromagnetic tendencies are
naturally also reduced. 
The computational study shows that a regime with sharp spin
incommensurate characteristics dominates in this region.
The orbital order remains staggered and the charge is uniformly spread.
This state is not directly related with the goals of the paper, 
and thus, the discussion of its origin and characteristics is
postponed for future work.

\subsubsection{Case of $n$$\ne$1}

One of the main goals of the study in this section is the
investigation of whether PS tendencies appear in a multi-orbital model
having only Coulomb interactions, starting at $n$=1 in the FM/OAF
regime previously observed with JT-phonons. 
For this purpose, here the couplings were fixed to $U'$=30 and $J$=22,
i.e., inside the FM/OAF phase of Fig.~7(b), and the density was varied 
using a cluster of 20 sites.
As the number of electrons $N$ was changed between 6 and 28 (only using
an even number), it was observed that the ferromagnetic
characteristics persist and $S(k)$ continued to be sharply peaked at
zero momentum.

\begin{figure}[h]
  \vskip-0.8truein
  \centerline{\epsfxsize=3truein \epsfbox{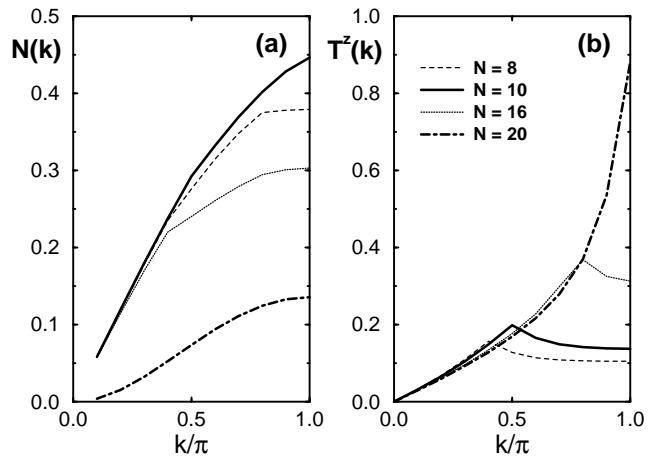} }
  \vskip-0.7truein
  \label{fig8}
  \caption{
    (a) DMRG result for the charge structure factor $N(k)$ vs. $k$ at 
    $n$=1 for $H_{\rm C}$ with the unit matrix hopping amplitude, 
    and a 20-site 1D chain.
    $U'$ and $J$ are fixed as 30 and 22, respectively, corresponding to
    couplings inside the FM/OAF phase in Fig.~7(b) . 
    The number of electrons $N$ is shown in the figure. 
    Note that the $k$=$\pi$ response is enhanced at density $n$=0.5, 
    at least relatively to the other densities.
    (b) Orbital structure factor $T^z(k)$ vs. $k$ using the same
    convention and couplings as in (a).
    Note the appearance of incommensurate characteristics in this
    channel as the density changes away from $n$=1.
    }
\end{figure}

Regarding charge-ordering, an enhancement in this channel was found at 
density $n$=0.5 as can be observed in Fig.~8(a) where results for
$N$=8, 10, 16, and 20 are presented.
This enhancement shows that at the couplings studied here tendencies
toward a CO-state are developed at $n$=0.5, in agreement with recent
MC studies for cooperative JT-phonons,\cite{hotta4}
the analysis of the previous sections with non-cooperative phonons,
and with experiments for manganites.\cite{tokura} 
The results in the orbital sector are shown in Fig.~8(b). 
As the density is reduced, or increased, starting at $n$=1 the orbital
correlations develop incommensurate characteristics, which is a
curious effect not observed before to the best of our knowledge.
When the density reaches $n=0.5$, $T^z(k)$ peaks at $k$=$\pi/2$,
effect likely correlated with the precursors of charge-ordering found
in $N(k)$.

\begin{figure}[h]
  \vskip-0.8truein
  \centerline{\epsfxsize=3truein \epsfbox{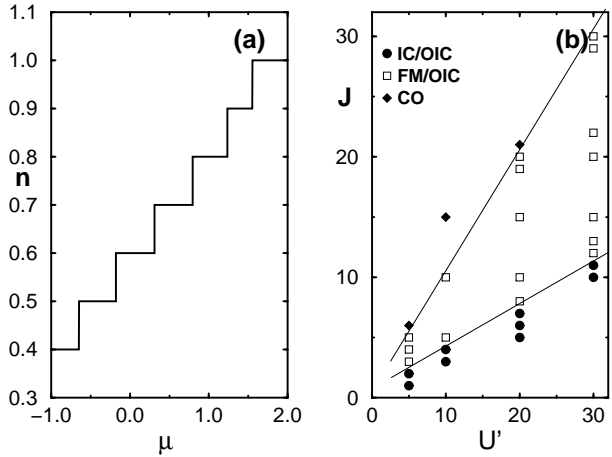} }
  \vskip-0.7truein
  \label{fig9}
\caption{
(a) Density $n$ vs. $\mu$ constructed from the ground-state energies 
corresponding to several number of electrons on a chain with 20 sites,
and the couplings used in Fig.~7. 
Note that all densities are accessible for some values of $\mu$.
(b) Phase diagram of $H_{\rm C}$ for the case of the unit hopping
matrix at $n$=0.5.
The two phases identified are IC/OIC and FM/OIC,
where ``OIC'' means charge-disordered orbital-incommensurate.
The charge-ordered regime above $J$=$U'$ has not been studied in as
much detail as in Fig.~7(b), and it is simply denoted by CO.
}
\end{figure}

Note that with the DMRG and ED techniques, which are setup in the 
canonical ensemble where the number of particles can be fixed 
arbitrarily, the stability of the various states with $N$-electrons
cannot be addressed directly.
For this purpose it is necessary to compare the energies of the
various ground states and construct a plot of the density $n$ vs. $\mu$ 
following steps already described in detail in previous
publications,\cite{yunoki2} and in the previous section.
Figure 9(a) shows that $all$ the densities studied here are actually
stable in the sense that a finite window of $\mu$ exists for all of
them where the state under study minimizes the energy.
This has to be contrasted with the results found in the MC
and mean-field calculations
with JT-phonons where states in a finite window of $n$ were found to
be unstable,\cite{yunoki4} i.e. there was no value of $\mu$ that
render them the global ground-state of the system. 
Then, it is concluded that the two-orbital Hubbard model with
unit-matrix hopping studied here does not phase-separate in 
spite of having characteristics at $n$=1 very similar to those
observed in the MC simulations of the JT-model at the same density.
Having a FM/OAF state at $n$=1 apparently is not sufficient for
PS to occur.\cite{imada}
This issue is conceptually important and will be addressed again in
the next subsection when results with a more realistic nondiagonal
hopping matrix are analyzed.

Due to the stability of the intermediate phases away from $n$=1 it
is not too surprising to observe a phase diagram at $n$=0.5 with
similar characteristics to those found in Fig.~7(b).
In Fig.~9(b), the unphysical region $J$$>$$U'$ has only been 
analyzed briefly, just sufficiently to confirm that CO characteristics exist
there, with relevant momentum $k$=$\pi/2$. 
In the regime $J$$<$$U'$, the spin FM and incommensurate phases are 
still stable, and $T^z(k)$ presents a broad peak at $k$=$\pi/2$
compatible with the tendencies toward charge-ordering that appear at
$n$=0.5.
The probable pattern of orbitals here may have orbital ``a'' 
below ``b'' at site i, a mostly empty i+1 site, the reverse orbital
pattern at i+2, and another empty site at i+3, with this arrangement
repeated in space.

\subsection{Nondiagonal Hopping Matrix}

The unit hopping matrix used in the previous subsection is a simple
and natural choice to gather qualitative information about the
two-orbital problem. 
However, the studies of models designed for manganites need a more
complicated hopping matrix, as shown in Sec.~III.
In this subsection, the realistic hopping amplitudes along the
$y$-direction are used to contrast the results against those obtained 
with the unit hopping matrix.
The results are actually found to be dramatically different at densities
different from unity.
The organization of the subsection is similar as the previous one,
i.e., the analysis starts with density $n$=1, establishing the main
features of the phase diagram, and continues with $n$$\ne$1 with
emphasis on the possible appearance of PS.

\subsubsection{Case of $n$=1}

In Fig.~10(a), the spin structure factor $S(k)$ is shown at $U'$=30
for representative values of $J$. 
At $J$=10 and smaller, the signal was found to be clearly
antiferromagnetic with a sharp peak at $k$=$\pi$.
In the intermediate region, the spin structure is complex with some 
incommensurate characteristics, as exemplified by the result at $J$=12.
For larger values of $J$, such as 15 and 20, the system becomes
ferromagnetic or quasi-ferromagnetic. 
If $J$ is increased further beyond the $J$=$U'$ boundary, spin
incommensurate structures have been observed, as in the study leading 
to Fig.~7(b) (results in this unphysical regime will not be discussed 
further here).
In Fig.~10(b), the charge structure factor $N(k)$ is shown at the same
couplings used in Fig.~10(a).
This quantity only develops some nontrivial structure as the $J$=$U'$
line is approached. 
In this case indications of charge-ordering at $k$=$\pi$ are observed.
In Fig.~11(a), the results corresponding to the orbital structure
factor are shown. 
At small $J$, the order is uniform since $T^z(k)$ peaks at $k$=0. 
At $J$=12, an incommensurate structure appears in the orbital sector,
similar to the results obtained for the unit-matrix hopping away from
$n$=1, and related with the spin incommensurability observed in
$S(k)$. 
For $J$=15 and 20, a robust orbital-staggered pattern is reached.

\begin{figure}[h]
  \vskip-0.8truein
  \centerline{\epsfxsize=3truein \epsfbox{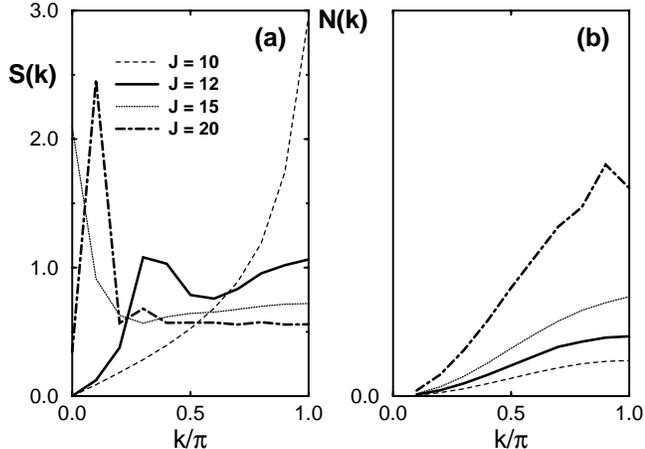} }
  \vskip-0.7truein
  \label{fig10}
  \caption{
    (a) Spin structure factor $S(k)$ vs. $k$ using the realistic hopping
    matrix along the $x$-axis for a 20-site 1D chain at $n$=1 and $U'$=30.
    The dashed, solid, dotted, and dotted-dashed lines correspond to
    $J$=10, 12, 15, and 20, respectively.
    (b) Charge structure factor $N(k)$ vs. $k$ for the same parameters 
    as in (a).}
\end{figure}

With the information obtained at $U'$=30 and various $J$'s,
supplemented by results gathered at $U'$=5, 10 and 20, a rough phase
diagram can be constructed, shown in Fig.~11(b). 
In this case four regimes are identified. 
At small $J$, an AF/OF state appears which does not exist for the
unit-matrix hopping. The reason is the following:
For the realistic hopping matrix, orbital ``a''  has more mobility
than the other. Then, to improve the kinetic energy the ground-state
prefers to have those orbitals as the lowest-energy ones at every site.
However, at $n$=1, a spin aligned orbital-uniform arrangement would
not have any mobility due to the Pauli principle. 
Then, the optimal situation is achieved with antiferromagnetic order
in the spin sector. 
The presence of this state is the main difference at $n$=1 between the
results found for the unit hopping matrix shown in Fig.~7(b) and the
results of Fig.~11(b).  
On the other hand, note that the AF/OF order does not take advantage
of $J$ since the use of the hoppings in this state leads to
antiparallel spins at the same site and, thus, to an energy
penalization proportional to $J$.
For this reason, as the coupling $J$ grows, a transition is expected
to a state in close competition with the AF/OF one, 
namely, the FM/OAF state that has appeared several times in our
investigations. 
In the latter $J$ plays an important role since it favors spins 
alignment when two electrons visit the same site, leading to a spin
ferromagnetic state. 
To optimize the kinetic energy, the optimal orbital pattern must be
staggered, as discussed in the previous section when explaining
Fig.~7(b).
The interpolation between these two competing states is difficult to
predict and the computational work suggests that it proceeds through a
complicated mixture of FM- and AF-orbital characteristics that lead to
an overall spin and orbital incommensurate pattern. 
Only further studies incorporating finite-size analysis will clarify
if this regime is indeed stable in the bulk limit. 

\begin{figure}[h]
  \vskip-0.8truein
  \centerline{\epsfxsize=3truein \epsfbox{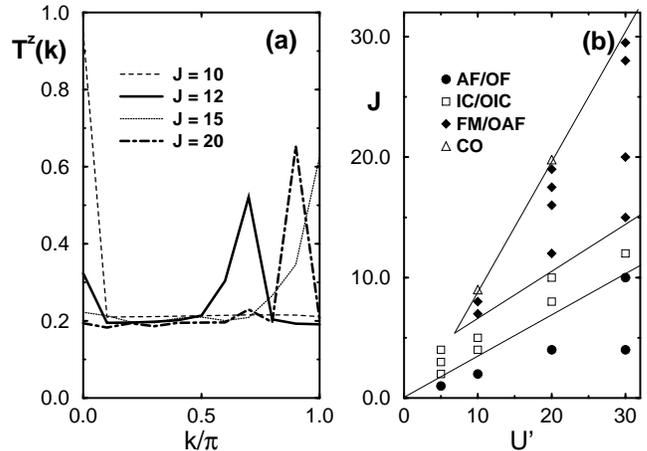} }
  \vskip-0.7truein
  \label{fig11}
  \caption{
    (a) Orbital structure factor $T^z(k)$ vs. $k$ for the same parameters
    as in Fig.~10(a).
    (b) Phase diagram for $n$=1 for the realistic hopping matrix.
    The four phases are AF/OF, IC/OIC, FM/OAF, and CO,
    where ``OF'' denotes orbital-uniform and charge-disordered.
    The ``CO'' has not been characterized in detail besides its CO
    properties.
    }
\end{figure}

\subsubsection{Case of $n$$\ne$1}

It is important to investigate how the phase diagram shown in Fig.~11(b)
evolves as a function of density. 
To establish a connection with the results found in the case of the
JT-phonons at large $\lambda$, once again the FM/OAF state is here
studied in detail since a state with similar characteristics was
indeed observed at $n$=1 (Sec.III) in the mean-field studies with phonons,
and the reduction of the density led to a phase-separated
regime in simulations.\cite{yunoki4} 
Would the same occur in the purely Coulombic case with non-diagonal
hopping?
To gain insight, $S(k)$ is shown in Fig.~12(a) at some representative 
densities using couplings $U'$=30 and $J$=15 that correspond to a
point inside the FM/OAF phase of Fig.~11(b).
It is interesting to observe an $abrupt$ change in $S(k)$ when $n$ is
varied from 1 to 0.9, with an incommensurate structure appearing in
the latter. 
This incommensuration continues up to $n$=0.5. 
The charge structure factor $N(k)$ also presents an abrupt change away
from $n$=1, with tendencies to charge ordering maximized at $n$=0.5
(not shown).
In addition, $T^z(k)$ drastically switches from an orbital-staggered 
pattern at $n$=1, to a uniform arrangement at $n$$\neq$1 as shown in 
Fig.~12(b). 
At $n$=0.5 and working with $U'$=10 and 30, it was observed that the
spin-incommensurate and orbital-uniform characteristics are
independent of $J$ to a good approximation, as long as $U'$$>$$J$.

\begin{figure}[h]
  \vskip-0.8truein
  \centerline{\epsfxsize=3truein \epsfbox{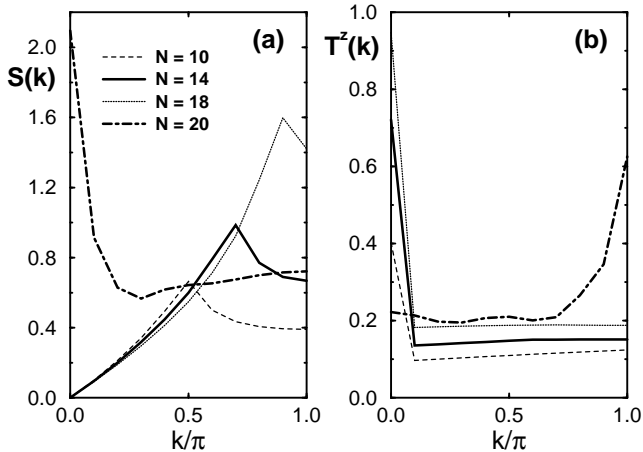} }
  \vskip-0.7truein
  \label{fig12}
  \caption{
    (a) Spin structure factor $S(k)$ vs. $k$ for the realistic hopping
    matrix, $U'$=30, and $J$=15 in the 20-site chain for several electron
    numbers.
    Incommensurate characteristics appear abruptly upon doping the $n$=1
    state.
    (b) Orbital structure factor $T^z(k)$ vs. $k$ for the same parameters
    and electron number as in (a). Again an abrupt change from OAF to OF
    features is observed as $n$ is reduced from unity.
    }
\end{figure}

To understand the curious behavior reported in Figs.~12(a) and (b),
the density vs. $\mu$ is plotted in Fig.~13(a) to search for
the PS tendency.
An anomalous behavior is observed at $n$=0.9, which has a tiny
$\mu$-window of stability, and thus, a large compressibility $\kappa$
(since $\kappa$ is proportional to d$n$/d$\mu$). 
This behavior indicates a tendency toward PS in the data, and it may
occur that small additions to the two-orbital Hubbard model such as
JT-phonons or even the analysis of slightly larger clusters, may
render the system truly phase-separated.
To confirm the presence of PS precursors, the local
density $n_i$ is shown in Fig.~13(b) at $n$=0.9.
Note that the DMRG method works with open boundary conditions, and
$n_i$ is not necessarily equal to $n$ at every site due to the lack of
translational invariance.
It is clear from this figure that large charge inhomogeneities are
present in the ground state, while at $n$=1 (not shown) $n_i$ is 
almost uniform. Only the center of the chain has a density equal to
the average one (0.9), while the ends of the chain have $n$ close to
unity, and at other sites near the center the density is smaller than
0.9. 
The minimum in the local density shown in Fig.~13(b) may correspond to
a hole doped into the $n$=1 system, that has developed polaronic
characteristics.
By symmetry, the other hole is on the other half of the chain. 
In Fig.~13(c), two holes generate two minima in the density (again,
with the other two holes on the other chain half). 
These results show that tendencies toward charge inhomogeneities
are present in this system, and precursors of PS are observed possibly
in the form of polaronic behavior. The purely Coulombic model appears
to be at the verge of phase separation.

\begin{figure}[h]
  \vskip-0.8truein
  \centerline{\epsfxsize=3truein \epsfbox{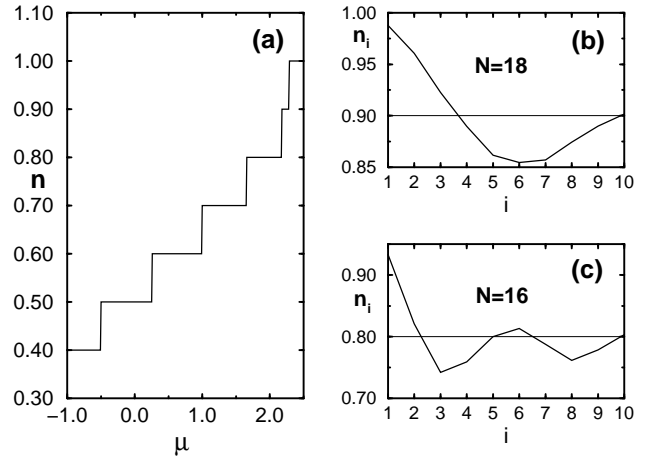} }
  \vskip-0.7truein
  \label{fig13}
\caption{
(a) Density $n$ vs. $\mu$ for the same parameter as in Fig.~12(a).
Precursors of PS at density close to unity are observed,
as discussed in the text.
(b) Local density $n_{i}$ vs. site position i for 18 electrons on a 20
site chain. Only half the chain is shown, the results for the other
half are simply obtained by reflection. The system is far from uniform
in the charge sector.
(c) Same as (b) but for 16 electrons.
}
\end{figure}
To search for more clear indications of phase-separation, the 
Coulombic interaction $U'$  was reduced to 10, and $J$ was fixed
to 7. This still corresponds to a point at $n$=1 inside the spin-FM
orbital-staggered phase.  
The spin and orbital structure factors for several number of electrons 
were calculated in this case, and the results (not shown) 
have clear similarities with those obtained at $U'$=30.
However, in this case now the analysis of the density vs
$\mu$ reveals strong phase separation characteristics between 
densities 1.0 and 0.7 (Fig.~14 (a)), 
qualitatively similar to those reported using JT-phonons.
In Figs.~14 (b) and (c), the local density away from $n$=1 is
shown. As in the case reported 
in Figs.~13 (b) and (c), strong oscillations reveal clear tendencies
to phase-separation. 
Studies at $U'$=20 and $J$=12 but with the realistic hopping matrix 
have also been carried out as part of this effort. The results are
very similar to those shown in Figs.~13 and 14.
Summarizing, either clear phase separation or a strong tendency to 
such phenomenon exists in the 1D purely Coulombic model as long as
the hopping matrix is non-diagonal.

\subsection{Influence of Nearest-Neighbor Repulsion}

As discussed in Sec.~IV, $V$ should not be too large in the actual
material, since if it were strong, the CS structure would be destroyed.
However, it is important to understand whether the results change or
not with the inclusion of $V$ using the same unbiased technique as in 
the previous subsection.

\begin{figure}[h]
  \vskip-0.8truein
  \centerline{\epsfxsize=3truein \epsfbox{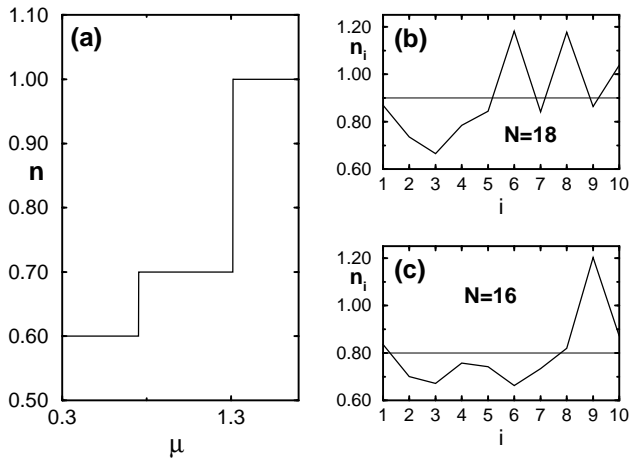} }
  \vskip-0.7truein
\caption{
(a) Density $n$ vs. $\mu$ for $U'$=10 and J=7. Clear phase separation
between densities 1.0 and 0.7 is observed.
(b) Local density $n_{i}$ vs. site position i for 18 electrons on a 20
site chain. Only half the chain is shown, the results for the other
half are simply obtained by reflection. The system is far from uniform
in the charge sector.
(c) Same as (b) but for 16 electrons.
}
\end{figure}

\begin{figure}[h]
  \vskip-0.8truein
  \centerline{\epsfxsize=3truein \epsfbox{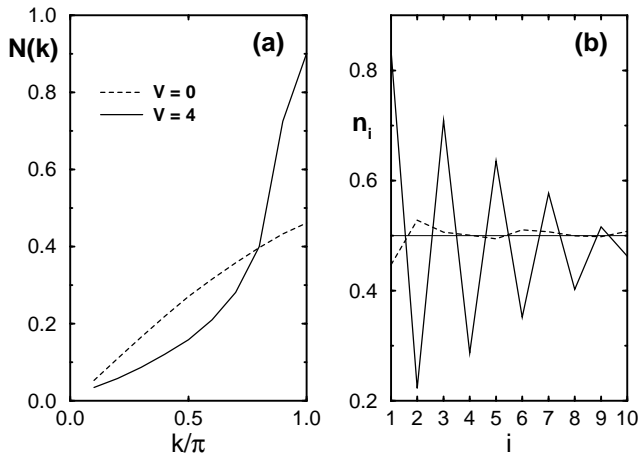} }
  \vskip-0.7truein
\caption{
(a) Charge structure factor $N(k)$ vs. $k$ at $n$=0.5 with the
realistic hopping for $U'$=10 and $J$=6 in the 20-site chain.
Two values of $V$ as shown and the enhancement of charge-order is
clearly observed.
(b) The local density $n_i$ vs. i for the same parametre as in (a).
Shown are results for half the lattice, the other half can be found by 
reflection.
}
\end{figure}

First let us consider density $n$=1, on a chain of 20 sites and
$U'$=30. In this case, few modifications were 
observed compared with the results obtained for $V$=0. 
Namely, (i) the boundary between the spin-AFM and spin-incommensurate
phases at small $J$, compared with $U'$, shifted toward a smaller $J$,
and (ii) close to $J$=$U'$ the charge correlations were enhanced 
substantially favoring the staggered pattern of charge between $n_i$=0
and 2, which is not penalized by $V$.
Even at values such as $J$=20, i.e. not too close to $J$=$U'$, this
enhancement was noticeable. 
However, since there are no indications of such a pattern in
experiments for manganites at $n$=1, there is no need to analyze this 
region in more detail in the present study.

At $n$=0.5, it is naively expected that the $V$-term brings the CO
state, although only precursors of CO behavior were observed at $V$=0
and this density as discussed in the previous subsection.
In Fig.~15(a), $N(k)$ is shown at $U'$=10 and $J$=6 both at $V$=0 
and 4. 
The region near $k$=$\pi$ is clearly enhanced by $V$ as expected.
The real-space density is shown in Fig.~15(b) and a clear
charge-staggered pattern is visible.
The spin structure factor (not shown) is still peaked at $k$=$\pi/2$
as at $V$=0, compatible with a staggered spin-arrangement involving
the occupied (even or odd) sites.
The orbital structure factor still has a large uniform component,
although it has also developed a broad low-intensity peak at
$k$=$\pi$.
Overall the characteristics of the state stabilized by $V$ are
orbital-uniform, charge-staggered with period of two lattice spacings, 
and spin-staggered over the occupied sites (period of four lattice
spacings).
Thus, only when a sufficiently large nearest-neighbor repulsion is
included, the $n$=0.5 CO-state similar to that found in the JT-phononic
model is reproduced in the purely Coulombic model in one dimension.
This result seems in contradiction with the discussion around the
CS structure observed in the
half-doped material, since it was concluded that $V$ is not 
the only origin of the CO-state in the manganite.
However, as briefly discussed in Sec.~III C, if the zigzag 1D chain,
not the straight 1D chain, is used in the calculation, the CO phase
can be obtained even in the purely Coulombic model without $V$,
although the $(3x^2-r^2)/(3y^2-r^2)$-type OO-phase 
cannot be obtained. From this viewpoint, a DMRG study carried out on 
the zigzag 1D chain will be interesting, but this is left for
the future.

The analysis of the two-orbital model that contains only Coulomb
interactions was instructive in several respects.
For instance, the study has shown that a spin-ferromagnetic
orbital-staggered (FM/OAF) phase appears naturally in this context 
at $n$=1 for a variety of hopping amplitudes, result in excellent
agreement with those observed for a purely JT model, also in agreement
with the previous ED studies of Coulombic models.\cite{rts}
Then, the FM/OAF phase is a robust feature of models for manganites
and approximations that attempt to describe these materials $cannot$
neglect orbital-ordering.
Another important result of this section has been the observation 
of PS tendencies with Coulombic-only interactions. 
This was shown to occur in the form of robust precursors of this
tendency in some regions of parameter space.
However, to observe PS 
it seems necessary to use realistic electronic hopping 
amplitudes in the case of Coulombic models. 
The presence of PS tendencies
both using purely JT and Coulombic interactions
confirms the robustness of this feature, and it also shows that
mixed-phase tendencies cannot be ignored in theoretical studies of
manganites.
In addition, the similarities between models with JT-phonons or
Coulombic repulsions suggest that the technically much simpler studies 
that use only phonons to mediate the interaction between electrons are
qualitatively correct, and likely capture the physics of more involved
models where both interactions are included.

\section{Discussion and Summary}

In this paper, the two-orbital model with the JT-phononic
and/or the Coulombic interactions has been studied using a variety of
techniques.
Three points have been confirmed in this work.
(i) The main properties of manganites can be reproduced successfully 
by the purely JT-phononic model even if the strong on-site correlation
is not included explicitly, since the effect of such correlation can be
renormalized into the effective electron-phonon coupling.
(ii) In particular, in the mean-field level,
the JT-phononic model can successfully reproduce the $CE$-type
AFM phase with the charge-stacked structure of the 3D cubic lattice 
at $n$=0.5, in excellent agreement with the MC simulations of 
Yunoki {\it et al.}
Even if the nearest-neighbor repulsion $V$ is introduced, this phase
is not easily destroyed due to the key role played by the magnetic
energy gain, regulated by $J_{\rm AF}$, in the $CE$- vs. s$CE$-type
competition and the ``topological'' energy gain in the $CE$-
vs. $C$-type competition.
Particularly, it is stressed that the topology of the zigzag 1D path
is the key issue leading to the stabilization of 
the CO/OO state in the $CE$-type structure. The on-site Coulombic
model treated in the mean-field approximation and without JT phonons
was also found to lead to charge-stacking due to the influence of
$J_{\rm AF}$.
(iii) The purely Coulombic model behaves in many respects very
similarly to the purely JT-phononic one and, in particular, it
presents the phase separation tendency, especially when realistic
hoppings are used. 

Summarizing, both approaches to the problem of manganites, based
either on Coulomb repulsions or phonons, share common tendencies.
This conclusion is in agreement with the recent
observation\cite{disorder}
that the percolative character of transitions in manganites and its
large magnetoresistance effect arise from the competition between
metallic and insulating phases in the presence of disorder,
independently on whether these phases are mainly generated by
Coulombic or JT interactions. 
Our results have provided robust arguments suggesting that
perceiving the ``Coulombic'' and ``JT-phononic'' approaches to
manganites as qualitatively different ways to carry out theoretical
calculations is likely incorrect. 

\acknowledgments

The authors thank S. Yunoki, A. Moreo, and S. Kivelson for
useful conversations.
T.H. has been supported by the Ministry of Education,
Science, Sports, and Culture of Japan during his stay in the 
National High Magnetic Field Laboratory, Florida State University.
A.L.M. acknowledges the financial support from Funda\c c\~ao
de Amparo \`a Pesquisa do Estado de S\~ao Paulo (FAPESP-Brazil).
E.D. is supported by grant NSF-DMR-9814350.

\appendix
\section{Relation with Phase-Separation Theories for Cuprates}

Since a substantial portion of the paper is devoted to the issue
of phase separation in models with Coulomb interactions, 
in this Appendix extra comments are provided regarding PS 
in models for high-temperature superconductors (HTSC), 
issue which has been
under discussion for almost a decade.\cite{kivelson,kivelson2}.
In particular, it is important to clarify the relation between HTSC
phase separation and the PS phenomenon discussed here for manganites.
In fact, some authors strongly believe that any doped correlated
insulator should have PS, and in this respect the results for
manganite models would be a mere particular case of a more general 
framework.\cite{kivelson}
However, it has to be discussed in detail to what extend there is 
convincing theoretical evidence that indeed any doped insulator phase
separates. If this is not clear, the relation between PS in cuprates
and manganites weakens considerably. 
The discussion in this context also has to involve two other
important aspects of the problem, namely the microscopic origin of PS,
and its phenomenological consequences, particularly when extended
Coulomb interactions are included.

There are several differences between the PS phenomena proposed for
manganites and cuprates.
{\bf (i)} The properties of the two competing phases are not the
same. In manganites, an undoped AFM-phase and a hole-doped FM-state
are involved, while in cuprates it is an undoped AFM-state and a 
hole-doped paramagnet or superconductor.
In models for HTSC, ferromagnetism does not play an active role at
realistic $J/t$ couplings, while it is crucial in manganites.
Nevertheless in Ref.\onlinecite{kivelson2}, PS between a G-type
antiferromagnet and a ferromagnet was also discussed, within the
framework of the very small $J/t$ limit of the $t$-$J$ model and
assuming a fully saturated FM-state. 
The argument in Ref.\onlinecite{kivelson2} can be applied directly to 
the case of the one-orbital model for manganites to justify the
presence of PS in this context. 
However, note that the AFM-state that is doped in 3D real manganites
has staggered spin order only in $one$ direction, while it is
ferromagnetic in the other two ($A$-type AFM order).
In fact, PS can occur in two-orbitals 
models for manganites in 1D and 2D 
without actually involving the AFM-state, but only having
two competing FM-states, as discussed in the main text.\cite{yunoki4} 
Here the orbital degree of freedom and JT phonons play the key role
needed for PS, while they are not important in the cuprates.
{\bf (ii)} PS in models for manganites occur even in one-dimension in
regions of couplings that are realistic, as described in this paper
and previous ones, unlike $t$-$J$ model results in the same dimension
where PS only happens at large values of $J/t$, and it does not occur
at all in the one-band Hubbard model.
{\bf (iii)} The study of PS in models analyzed in the infinite
dimensional limit lead to different conclusions between cuprates and
manganites, although this issue is still controversial.
Fair is to say that the case of the $D$=$\infty$ Hubbard model is
subtle and deserves a special discussion ($D$ is dimensionality). 
In principle at $D$=$\infty$ there are no indications of PS in the
one-band Hubbard model.\cite{georges}
However, it is not clear if this can be taken as a counterexample 
of PS in a Mott insulator, as recently remarked in 
Ref.\onlinecite{carlson}.
In most $D$=$\infty$ investigations, the ratio $J$/$t$ scales to zero
like 1/$\sqrt{D}$ and as a consequence the physics of PS is difficult
to address.
In addition, when non-bipartite lattices are used, the AFM order is
frustrated. 
Then, more work is needed in large dimension to clarify the effect of
doping on correlated insulators.
However, note that in calculations for the one-orbital Kondo model for
manganites PS has been clearly observed even at $D$=$\infty$ (see 
Ref.\onlinecite{yunoki}). 
This suggest that important qualitative differences may exist between
the PS phenomena in models for cuprates and manganites.

Let us elaborate more on stripes in manganites and 
its topological vs. non-topological character.
While stripe order has been claimed to exist in 
manganites,\cite{mori} note that the regime of hole density in 
which this observation was made ($x$=2/3) is far from the low hole
concentration were the discussion of stripes in the cuprates occurs.
These authors are not aware of experiments reporting stripes at small
$x$ in manganites, although charge inhomogeneities of various forms
appear in several cases in this regime.\cite{moreo}
In addition, it is unclear whether the bi-stripe structure found by
Mori {\it et al.}\cite{mori} has any topological characteristics.
Recent calculations by Hotta {\it et al}.\cite{hotta2} have shown that 
the manganite bi-stripes are better interpreted as arising from zigzag
conducting chains running $perpendicular$ to the charge stripes
once electron-phonon JT-couplings are switched-on.
For this reason, it is premature to establish connections between
stripes in cuprates and manganites.

Regarding the issue of whether any doped correlated insulator produces
PS, the following is our understanding of the current theoretical
literature for the cuprates.
The most clear manifestation of PS appears in the $t$-$J$ model at
large $J/t$, where the tendency to form pairs of holes to minimize the
number of broken AFM bonds is so intense that clusters of holes are
formed instead of individual pairs.
Here the attractive potential energy among hole carriers originating
in the AFM background dominates over the kinetic energy, that tends to
spread particles apart. 
As $J/t$ is reduced it has been a matter of much controversy whether
the PS effect survives in the realistic small $J/t$ regime of the
$t$-$J$ model. 
While ED of small clusters,\cite{dagotto} 
MC simulations,\cite{moreo3,furu} and 
high-temperature expansions\cite{puttika} 
suggested that at $J/t$ of order unity the effect would disappear,
other arguments\cite{kivelson} and 
further computationalwork\cite{hellberg} opened the possibility for PS
to exist at all values of $J/t$.
Very recent DMRG studies\cite{white2} for ladders of increasing number 
of legs show that previous  calculations\cite{hellberg} may not have
been sufficiently accurate and the new results suggest that PS indeed
only occurs at intermediate and large values of $J/t$ in the 2D
$t$-$J$ model.
This same conclusion becomes more clear once extra hopping amplitudes
are added to the model.
In this case the substantial hole mobility induced by the extra
hoppings shifts the PS regime to values of $J/t$ larger than in the
pure $t$-$J$ model case.\cite{tohyama}
Then, the proposal that any correlated insulator when hole doped
should become phase separated is still not confirmed using unbiased
techniques in simple models for cuprates.

Regarding the phenomenological aspects of the PS regime, the issue of
microscopic PS that may appear in manganites has certainly been
discussed before by Emery and Kivelson for cuprates.\cite{kivelson} 
In this context the Coulomb interactions break into small pieces the
macroscopic clusters of the two phases in competition, since they have
different electronic densities. 
Stripe patterns emerged from calculations carried out mainly close to
the atomic limit,\cite{low} where the attraction leading to PS plus
the Coulomb repulsion are in competition.
In the context of Nuclear Physics similar patterns have also been 
discussed.\cite{nuclear}
In addition, Nagaev studied the formation of finite size clusters of
one phase embedded into the other, mainly for antiferromagnetic
semiconductors.\cite{nagaev}
Then, the simple picture of a $stable$ state formed by small clusters
of the competing phases, somewhat similar to the CDW pattern obtained
in the 1D calculations with nearest-neighbor repulsions, is certainly
common to manganites and cuprates, and it has been described in the
context of the Frustrated PS scenario for HTSC.\cite{kivelson}

In short, the discussion presented in this Appendix suggests that
the PS phenomena in models for manganites and cuprates, while sharing
the general common ingredients of any PS regime, are different in
origin and they must be considered separately in their study.




\end{multicols}

\begin{references}
\bibitem{comment3}
Note that this statement is shown to be valid in the limit of 
$J_{\rm H}$=$\infty$. 
For finte $J_{\rm H}$, it is an open problem whether the MFA becomes 
exact in the 1D case or not.
\bibitem{comment1}
In the one-orbital model, the 1D system will be insulating if the
electron-lattice coupling is non-zero.
In the two-orbital case, however, the situation is not so simple,
and the metal-insulator transition point may depend on the hopping
matrix. 
However, at least for the realistic hopping amplitude, the 1D system
is insulating for a finite electron-phonon coupling.
\bibitem{comment2}
It may be reasonable to interpret that the JT splitting observed in
the photoemission experiment is related to ${\tilde E}_{\rm JT}$, not
to the bare value $E_{\rm JT}$.
If the purely JT-phononic model is used to investigate the manganite,
$E_{\rm JT}$ appearing in the Hamiltonian may be better to be
interpretted as ${\tilde E}_{\rm JT}$.
In the present work, however, in order to focus on the competition
between the JT-phononic and Coulombic interactions, 
$E_{\rm JT}$ and $U'$ are treated separately.
As for the stabilization of the $CE$-type AFM state with CS structure,
values of $E_{\rm JT}$ and $U'$ are not so important. 
Rather, ${\rm E_{\rm JT}}$ large compared to $t$ is essential.
\end{references}
\end{document}